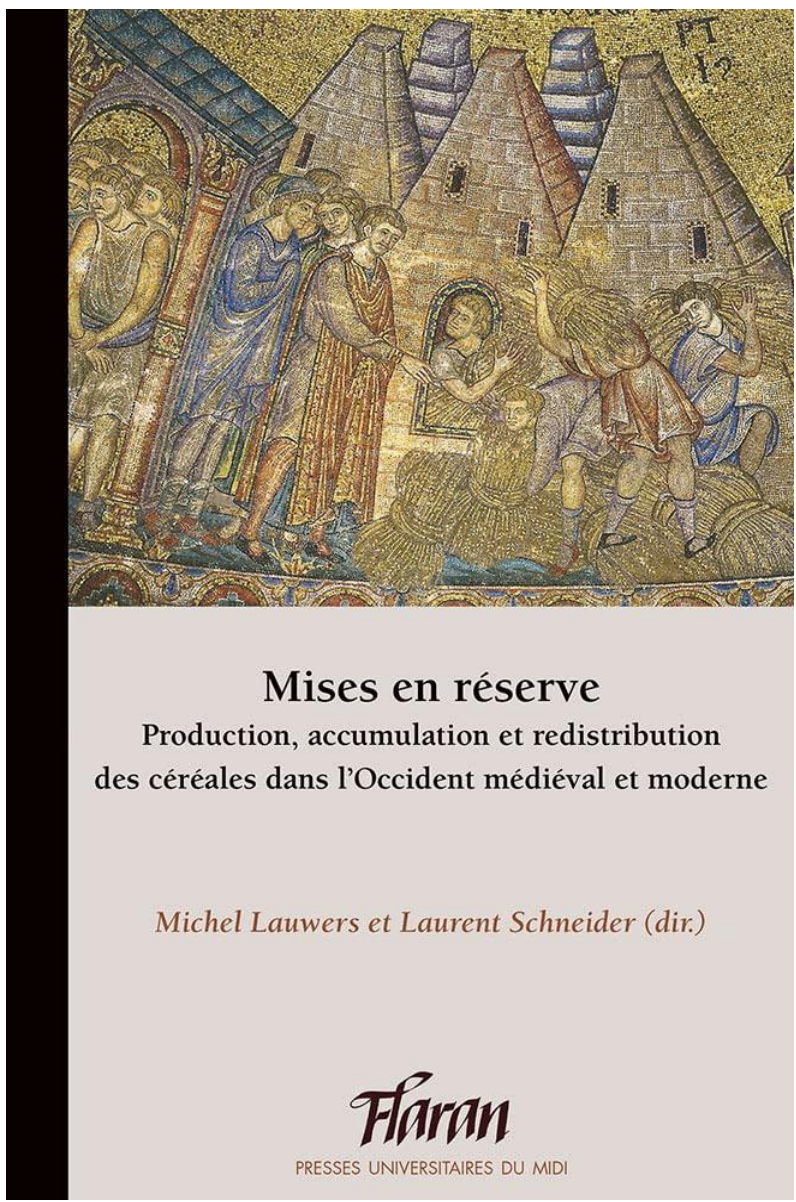

La pagination de l'article imprimé est indiquée entre crochets [xxx]

# Prélèvements, organisation spatio-temporelle et domination sociale.
# Les mentions de lieux de stockage dans les chartes médiévales (VIIe-XIIIe siècle)[1]

Nicolas Perreaux*

[143] « La conservation est une question technique, le stockage, une question économique ; on peut connaître les techniques sans vouloir les appliquer »[2]. Par cette phrase, Alain Testart revenait en 2012 sur un thème qui lui était cher, celui de la production des inégalités dans les sociétés[3]. [144] L'anthropologue voyait dans l'usage des greniers un vecteur de la « différentiation sociale »[4], à travers l'accumulation matérielle et ostentatoire que ceux-ci engendraient. La mise en réserve se situe en effet à l'intersection de différentes dimensions sociales et représentationnelles fondamentales. Elle implique une pensée originale du temps (puisqu'elle est liée à la saisonnalité des cultures, donc à une forme de prévision), mais aussi de l'espace (parce que c'est la sédentarité qui détermine la nécessité, ou non, de recourir massivement aux pratiques de stockage). Quasi-inévitablement, l'existence de greniers engendre une hiérarchisation au sein

---

* Nicolas Perreaux, Université de Paris 1 Panthéon-Sorbonne – LaMOP.
[1] Nous tenons à remercier vivement Hélène Débax, Julien Demade, Jean-Pierre Devroey, Alain Guerreau, Thomas Labbé, Michel Lauwers, Sandrine Lavaud, Jean-François Nieus, Édith Peytremann, Christian Sapin, Nicolas Schroeder et Roland Viader, pour leurs précieux conseils lors de la préparation de ce texte.
[2] A. Testart, *Avant l'histoire. L'évolution des sociétés de Lascaux à Carnac*, Paris, 2012, p. 206. L'auteur écrit encore : « [N]ul doute que les chasseurs-cueilleurs, même mobiles, connaissent maintes techniques de conservation, mais c'est une autre question de constituer des réserves pour tout l'hiver » (p. 330).
[3] A. Testart, *Les chasseurs-cueilleurs ou l'origine des inégalités*, Paris, 1982. La thèse de l'auteur consiste à faire du stockage une dynamique fondamentale de la différenciation des sociétés humaines, en dépassant l'opposition classique entre chasseurs-cueilleurs et sédentaires-agriculteurs. Il insiste en particulier sur la quantité du stock, selon lui plus déterminante dans les évolutions sociales que la connaissance des méthodes de stockage – souvent maîtrisées dans les sociétés, y compris celles composées de chasseurs-cueilleurs (stockeurs ou non).
[4] A. Testart, *Avant l'histoire…*, p. 208.

des communautés humaines, parce qu'ils accentuent les inégalités entre ceux qui peuvent en assumer la construction, qui en contrôlent le contenu ou l'accès et ceux qui en usent par nécessité – parfois sous la contrainte – sans en jouir pleinement[5]. Le stockage se situe donc au cœur des sociétés qui choisissent d'y recourir, tant pour la production agraire que pour la reproduction des rapports sociaux. Dans le cas de l'Europe médiévale, les implications du développement des espaces de stockage, ainsi d'ailleurs que les phases de celui-ci restent toutefois encore largement à déterminer.

Pour la période allant du VII[e] au XIII[e] siècle, ce sont d'abord les historiens des techniques qui se sont intéressés, dès les années 1970, aux pratiques de stockage[6]. Progressivement, les archéologues ont apporté des séries d'observations inédites, particulièrement autour des fosses-silos, mais aussi à travers l'archéologie expérimentale, l'étude des granges et la carpologie[7]. Ces avancées dans la connaissance des infrastructures [145] de stockage, en partie liées au développement de l'archéologie préventive, ont récemment permis la production de synthèses intermédiaires pour le haut Moyen Âge[8]. Depuis 2012, différents colloques ont participé au regain d'intérêt autour des lieux de stockage médiévaux[9].

---

[5] M. Sahlins, *Âge de pierre, âge d'abondance. L'économie des sociétés primitives*, Paris, 1976 ; J.C. Scott, Homo domesticus. *Une histoire profonde des premiers États*, Paris, 2019. Concernant l'apport anthropologique, voir le rapport de Michel Lauwers dans le présent volume.

[6] F. Sigaut, M.-C. Aubin, *Les techniques de conservation des grains et leurs fonctions sociales. Recherche de technologie culturelle*, Paris, 1979 ; M. Gast, F. Sigaut (dir.), *Les techniques de conservation des grains à long terme : leur rôle dans la dynamique des systèmes de cultures et des sociétés*, Paris, 1979-1985.

[7] Avec une attention portée sur les fosses-silos dès le tournant des années 1970-1980 (voir le rapport de Laurent Schneider dans le présent volume). Quelques travaux récents : É. Faure-Boucharlat (dir.), *Vivre à la campagne au Moyen Âge. L'habitat rural du V[e] au XII[e] siècle (Bresse, Lyonnais, Dauphiné) d'après les données archéologiques*, Lyon, 2001 (notamment les chapitres V et VI : « Les constructions rurales », « Vivre et travailler à la campagne », p. 77-126) ; L. Schneider, « Vers la Méditerranée : un regard sur les grandes aires d'ensilage médiévales », O. Maufras (éd.), *Habitats, nécropoles et paysages dans la moyenne et la basse vallée du Rhône (VII[e]-XV[e] siècle)*, Paris, 2006, p. 440-445 ; F. Gentili, « Silos et greniers : structures de conservation des grains sur les sites ruraux du haut Moyen Âge d'après des exemples franciliens », F. Gentili, A. Lefèvre, N. Mahé (dir.), *L'habitat rural du haut Moyen Âge en Île-de-France*, Guiry-en-Vexin, 2009, vol. 2, p. 97-123 ; L. Schneider, « Structures du peuplement et formes de l'habitat dans les campagnes du sud-est de la France de l'Antiquité au Moyen Âge (IV[e]-VIII[e] siècle). Essai de synthèse », *Gallia*, t. 64, 2007, p. 11-56 ; É. Peytremann, « Structures et espaces de stockage dans les villages alto-médiévaux (VI[e]-XII[e] siècle) de la moitié septentrionale de la Gaule : un apport à l'étude socio-économique du monde rural », A. Vigil-Escalera Guirado, G. Bianchi, J.A. Quirós (dir.), *Horrea, Barns and Silos. Storage and Incomes in Early Medieval Europe*, Bilbao, 2013, p. 39-56 ; C. Ebanista, « La conservazione del grano nel medioevo: testimonianze archeologiche », G. Archetti (dir.), *La civiltà del pane. Storia, tecniche e simboli dal Mediterraneo all'Atlantico*, Spolète, 2015, p. 469-521 ; J. Ros, M.-P. Ruas, L. Bouby, C. Hallavant, N. Rovira, E. Roux, « Un demi-siècle de carpologie : bilan des travaux dans les Pyrénées-Orientales (1969-2017) », *Archéo 66*, vol. 32, 2017, p. 85-97 ; J.-M. Poisson, « Espaces et modes de stockage des denrées alimentaires dans les maisons villageoises médiévales », F. Journot (dir.), *Pour une archéologie indisciplinée. Réflexions croisées autour de Joëlle Burnouf*, Drémil-Lagage, 2018, p. 235-245.

[8] Voir les articles d'Édith Peytremann et de Laurent Schneider mentionnés dans la note précédente.

[9] K.A. Bergsvik, R. Skeates (dir.), *Caves in Context. The Cultural Significance of Caves and Rockshelters in Europe*, Oxford-Oakville, 2012 ; Vigil-Escalera Guirado, G. Bianchi, J.A. Quirós

Dans cette tendance historiographique favorable, l'analyse des textes a toutefois été en partie délaissée, pour des raisons à la fois pratiques et disciplinaires[10]. L'analyse des mentions textuelles de silos ou de greniers pose en effet plusieurs difficultés : quels termes désignaient ces lieux ? Comment les repérer au sein d'un corpus remarquablement dense ? Enfin, comment faire face à leur nombre qui, à l'échelle européenne, est probablement remarquable ? Déterminer le rôle qu'ont pu jouer ces espaces de stockage [146] implique de connaître en effet, en premier lieu, leur fréquence et leur distribution chrono-géographique.

L'objectif de cet article est de réaliser un examen systématique des mentions de lieux de stockage de grains dans les chartes[11]. Dans un premier temps, on s'attachera à définir les contours de ce lexique, en proposant une liste de termes à explorer via différentes méthodes qualitatives et quantitatives. Il s'agira ensuite de définir la répartition chronologique de ces mentions, depuis le VII[e] jusqu'à la fin du XIII[e] siècle. Enfin, on tentera de cerner les évolutions sémantiques et contextuelles autour de ces lieux, afin d'éclairer les causes et les implications de la mise en réserve médiévale.

**Corpus, méthodes, lexiques**

*Terminologie et corpus : établir une première liste lexicale*

Cette enquête est aujourd'hui possible grâce aux progrès des méthodes numériques. Depuis quelques années, ces dernières ouvrent des perspectives complémentaires en médiévistique, en particulier pour les chartes qui ont été numérisées en masse lors de programmes nationaux et internationaux[12]. Depuis

---

(dir.), *Horrea, Barns and Silos. Storage and Incomes in Early Medieval Europe…* ; G. Archetti (dir.), *La civiltà del pane. Storia, tecniche e simboli dal Mediterraneo all'Atlantico…* ; C. Alix, L. Gaugain, A. Salamagne (dir.), *Caves et celliers au Moyen Âge et à l'époque moderne*, Rennes, 2019.
[10] Quelques exceptions notables : G. Comet, *Le paysan et son outil. Essai d'histoire technique des céréales (France, VIII[e]-XV[e] siècle)*, Rome, 1992 (coll. École française de Rome n° 165) ; J.-P. Devroey, J.-J. Van Mol (dir.), *L'épeautre, histoire et ethnologie*, Treignes, 1989 ; J.-P. Devroey, J.-J. Van Mol, C. Billen (dir.), *Le seigle (Secale cereale), histoire et ethnologie*, Treignes, 1995 ; J.-P. Devroey, A. McBride, « Food and Politics », M. Montanari (dir.), *A Cultural History of Food in the Medieval Age*, Londres-New York, 2012, vol. 2, p. 73-89 ; J.-P. Devroey, « La céréaliculture au haut Moyen Âge (III[e]-XI[e] siècles AD). Apports archéologiques et problèmes historiques récents », *Vie archéologique. Bulletin d'information trimestriel de la Fédération des archéologues de Wallonie*, vol. 72, 2013, p. 59-66 ; J. Claridge, J. Langdon, « Storage in Medieval England : The Evidence from Purveyance Accounts, 1295-1349 », *Economic History Review*, vol. 64-4, 2011, p. 1242-1265.
[11] Ce corpus possède différents avantages (sa relative homogénéité thématique et linguistique, ainsi que sa distribution à l'échelle européenne, sur la totalité de la période considérée), qui permettent un traitement diachronique des occurrences.
[12] Avec par exemple les programmes CBMA, *Chartae Galliae*, Artem, Fundació Noguera, Codice Diplomatico della Lombardia Medievale, Deeds Project, dMGH, Scripta, Württembergisches Urkundenbuch, Cartulaires d'Île-de-France, Monasterium et Diplomata Belgica. Voir A. Ambrosio, S. Barret et G. Vogeler (dir.), *Digital diplomatics. The computer as a tool for the diplomatist?*, Köln-Weimar-Wien, 2014 (*Archiv für Diplomatik, Schriftgeschichte, Siegel- und Wappenkunde*, Beiheft 14) ; N. Perreaux, « L'écriture du monde (résumé, parties I et II) », *Bulletin du Centre d'études médiévales d'Auxerre*, vol. 19.2 et vol. 20.1, 2016, <https://cem.revues.org/14264> et <https://cem.revues.org/14452>.

2009, nous nous sommes attachés à réunir ces éditions numérisées dans un corpus documentaire, les *Cartae Europae Medii Aevi* (CEMA). Il regroupe aujourd'hui plus de 225 000 documents lemmatisés[13].

Toute analyse lexicale dans les corpus numérisés débute nécessairement par l'établissement d'une liste de vocables assez fréquents pour être étudiés de façon systématique. Or, la variété des termes désignant des dispositifs de stockage paraît de prime abord assez déroutante. Une [147] recherche dans l'historiographie[14], les glossaires ou dictionnaires[15], et bien entendu dans la documentation diplomatique (*i.e.* les chartes) permet de dégager une quarantaine de termes, représentant au total plus de 15 000 occurrences dans les CEMA (cf. fig. 1). Ce nombre paraît élevé, mais il s'agit le comparer par exemple avec *molendinum*, qui regroupe 27 421 mentions, ou encore à *villa*, avec 77 272 occurrences[16]. On peut aussi rapprocher ces scores de ceux obtenus pour la *Patrologie latine*, où apparaissent 10 997 mentions de ce lexique, soit moins que dans les chartes, avec une répartition d'ailleurs assez différente[17]. [148]

---

[13] Le corpus a été formalisé pour différents logiciels permettant des manipulations textuelles, de la recherche simple à la fouille de données (TXM, CQB, R).

[14] W. Winckelmann, « Die Ausgrabungen in der frühmittelalterlichen Siedlung bei Warendorf (Westfalen) », W. Krämer, *Neue Ausgrabungen in Deutschland*, Berlin, 1958, p. 492-517 ; H. Dölling, *Haus und Hof in Westgermanischen Volkrechten*, Münster, 1980 ; É. Peytremann, « Structures et espaces de stockage dans les villages alto-médiévaux… », p. 42 ; L. Schneider, « Structures du peuplement et formes de l'habitat… ». Pour le latin antique, voir A. Ferdière, « Essai de typologie des greniers ruraux de Gaule du Nord », *Revue archéologique du Centre de la France*, t. 54, 2015, <http://journals.openedition.org/racf/2294> ; G. Huitorel, « Stocker les céréales dans les établissements ruraux du nord de la Gaule à l'époque romaine. Essai d'identification des modes de stockage entre le I[er] et le IV[e] siècle ap. J.-C. », *Les céréales dans le monde antique. Regards croisés sur les stratégies de gestion des cultures, de leur stockage et de leurs modes de consommation*, *Nehet, revue numérique d'égyptologie*, n° 5, 2017, p. 217-238 ; ici p. 222-223.

[15] En particulier le Du Cange en ligne, <http://ducange.enc.sorbonne.fr/>, ainsi que les volumes du *Novum Glossarium* disponible sur <http://glossaria.eu/>. L'équipe des projets ANR Omnia (dir. O. Guyotjeannin) et Velum (dir. B. Bon) a créé un dictionnaire médiolatin sous forme d'une base de données, où il est possible de faire des recherches inversées (par exemple : « greniers », « céréales », « stockage », etc.). Nous avons largement exploité cette dernière possibilité afin de récolter un maximum de termes.

[16] Pour l'analyse systématique de ce terme, nous nous permettons de renvoyer à N. Perreaux, « Chronologie, diffusion et environnement des *villae* dans l'Europe médiévale (VII[e]-XIII[e] siècles) : recherches sur les corpus diplomatiques numérisés », S. Bully, C. Sapin (dir.), *L'origine des sites monastiques : confrontation entre la terminologie des sources textuelles et les données archéologiques*, Dijon, 2016, <https://cem.revues.org/14476> (*Bulletin du Centre d'études médiévales d'Auxerre*, Hors-série n° 10).

[17] La *Patrologie latine* demeure le plus vaste corpus de textes médiévaux, si l'on inclut les Pères de l'Église à cette période, avec environ 100 millions de mots. Une version libre du corpus est désormais consultable et téléchargeable à l'adresse suivante : <http://mlat.uzh.ch>.

| Lemme | CEMA | PL | Antique |
|---|---|---|---|
| grangia | 6444 | 561 | 0 |
| cella | 3852 | 5104 | 210 |
| granica | 1398 | 9 | 0 |
| cellarium | 1362 | 626 | 5 |
| horreum | 622 | 1698 | 169 |
| organeum | 476 | 0 | 0 |
| granarium | 317 | 46 | 19 |
| canaba | 184 | 62 | 0 |
| cellarius | 200 | 76 | 4 |
| spicarium | 64 | 12 | 0 |
| cellaris | 55 | 181 | 1 |
| scuria | 55 | 26 | 0 |
| granatarius | 40 | 39 | 0 |
| cubiculum | 36 | 2148 | 357 |
| granea | 31 | 2 | 0 |
| graniam | 12 | 0 | 0 |
| promptuarium | 12 | 89 | 2 |
| apotheca | 12 | 252 | 9 |
| frumentarius | 8 | 50 | 149 |
| granaticum | 3 | 2 | 0 |
| paleare | 3 | 2 | 0 |
| boia | 0 | 0 | 0 |
| cellariensis | 0 | 0 | 3 |
| doliarium | 0 | 0 | 0 |
| farinaria | 0 | 0 | 0 |
| garberius | 0 | 0 | 0 |
| horreolanus | 0 | 0 | 0 |
| paracellarium | 0 | 1 | 0 |
| proma | 0 | 2 | 1 |
| salgamum | 0 | 3 | 3 |
| sirus | 0 | 0 | 3 |
| sitonicon | 0 | 6 | 0 |
| torbax | 0 | 0 | 0 |
| Somme | 15186 | 10997 | 935 |

**Fig. 1.** Mentions de structures de stockage, dans les CEMA (textes datés et non datés), la *Patrologie latine* (= PL) et un corpus antique (= Antique ; II[e] siècle a.C.-II[e] siècle p.C.). Surligné en gris : lemmes avec plus de 100 occurrences surreprésentés dans l'un des corpus[18].

[149] *Grangia* est en effet nettement plus présent dans les chartes que dans les autres corpus, tout simplement car il s'agit d'un terme absent de l'Antiquité païenne, mais aussi de la Vulgate, à l'inverse d'*horreum*[19] (fig. 1). Les espaces de stockage constituent donc un vocabulaire courant, sans toutefois appartenir au lexique le plus fréquent. Certains lemmes relevés dans les dictionnaires ou l'historiographie sont par ailleurs absents ou faiblement représentés dans les CEMA, par exemple *boia*, *doliarium*, *garberius*, *granaticum*, *horreolanus*, *paracellarium*, *proma*, *salgamum*, *sirus*, *sitonicon*, ou encore *torbax*. On trouve certes des mentions exceptionnelles pour certains termes, comme cette *aula* « [*qui*] *fuit plena de frumento infra postes usque ad trabes* », dans une charte-inventaire de

---

[18] Les termes avec « 0 » occurrence ont été relevés dans la littérature ou les dictionnaires. Nous les mentionnons pour comparaison.
[19] Sur *grangia* chez les cisterciens, voir D. Panfili, « *Domus*, *grangia*, *honor* et les autres. Désigner les pôles cisterciens en Languedoc et Gascogne orientale (1130–1220) », *Le Moyen Âge*, t. 123, 2017 (2), p. 311-338, ainsi que sa contribution dans ce volume. Concernant *horreum*, voir nos développements ci-dessous.

1229 pour l'archevêque de Rouen, Thibaud d'Amiens (1221-1229)[20]. Mais ces mentions ne permettent pas une enquête systématique. D'autres termes, parfois traduits par « grenier », « grange » ou encore « cellier », renvoient majoritairement à tout autre chose qu'au stockage[21]. C'est le cas par exemple de *cella*, avec plus de 3 800 occurrences dans les chartes, qui désigne principalement des structures monastiques, mais aussi de *camera*, *aula*, etc.[22].

Ce dernier cas montre toute la complexité, et donc l'intérêt, du programme scientifique exploré lors du colloque : les dispositifs de stockage constituent une catégorie fluide, depuis le coffre jusqu'au grenier[23]. L'existence d'une réserve, attestée par des fouilles ou par des mentions textuelles, ne garantit en outre pas automatiquement la présence de [150] céréales dans celle-ci[24]. Afin de déterminer une terminologie sur laquelle mener l'enquête, il s'agit donc de mesurer l'association des grains et du lexique désignant ces lieux dans les chartes.

---

[20] L. Delisle (éd.), *Études sur la condition de la classe agricole et l'état de l'agriculture en Normandie au Moyen Âge*, Évreux, 1851, n° 28 – Scripta n° 7493. L'acte-inventaire évoque l'état du manoir de Killon (Angleterre). Il contient d'autres descriptions intéressantes pour l'histoire du stockage des grains, par exemple : « *minor grangia fuit plena de avena ab occidentali parte usque ad postem ex altera parte una culaz fuit plena de avena* ».

[21] Dans cet article, nous avons retenu par convention « grenier » plutôt que « grange » afin d'évoquer les structures de stockage pérennes. Bien que l'étymologie des deux termes renvoie aux grains (respectivement *guernier*/*granarium* et *granica*), il nous semble que « grange » est légèrement plus ambigu et qu'il peut aussi désigner en français contemporain des lieux où se trouvent les bêtes (ce qui n'est pas le cas du « grenier »).

[22] *Aula* désigne par exemple fréquemment la cour royale ou impériale, mais encore une salle de réception seigneuriale et, par extension, une vaste pièce. Sur *cella*, voir M. Gaillard, « *Monasterium*, *cella*, *abbatia*… Enquête sur les différents termes désignant les communautés religieuses au haut Moyen Âge (V$^e$-milieu IX$^e$ siècle) et leur signification », S. Bully, C. Sapin (dir.), *L'origine des sites monastiques…*, <https://journals.openedition.org/cem/14474>

[23] Sur la variété des modes de stockage, voir É. Peytremann, « Structures et espaces de stockage dans les villages alto-médiévaux… », p. 39-42.

[24] Le colloque se concentrait exclusivement sur le stockage des grains. Il s'agirait de savoir dans quelle mesure cette activité impliquait des techniques, des lieux et des objets spécifiques. Le stockage des légumineuses, mais encore du foin, serait un autre champ à explorer, sans doute connecté à la question des céréales.

| Lemme | Occurrences | Associations | % |
|---|---|---|---|
| *granarium* | 317 | 85 | 26,81 |
| *spicarium* | 64 | 14 | 21,88 |
| *granea* | 31 | 6 | 19,35 |
| *horreum* | 622 | 79 | 12,7 |
| *granica* | 1384 | 143 | 10,33 |
| *granatarius* | 40 | 3 | 7,5 |
| *canaba* | 182 | 9 | 4,95 |
| *grangia* | 6435 | 318 | 4,94 |
| *cellarium* | 1362 | 20 | 1,47 |
| *cella* | 3306 | 1 | 0,03 |
| *organeum* | 476 | 0 | 0 |
| *scuria* | 53 | 0 | 0 |
| *cubiculum* | 36 | 0 | 0 |
| *graniam* | 12 | 0 | 0 |
| *apotheca* | 12 | 0 | 0 |

**Fig. 2.** Associations dans les CEMA entre les lieux de stockage et les mentions de grains (*siligo*, *hordeum*, *triticum*, *frumentum*, *civata*, *avena*, *spelta*, *granum*, *bladum*, etc.). « Occurrences » = mentions datées des lieux de stockage ; « Associations » : cooccurrences entre ces lieux de stockage et les mentions de grains (dans un contexte de plus ou moins cinq mots) ; « % » : pourcentage d'associations sur la totalité du lemme.

Partant des termes *siligo*, *hordeum*, *triticum*, *frumentum*, *civata*, *avena*, *spelta*, *granum*, *bladum*, etc.[25], il est alors possible de montrer que *grangia* (6 435 mentions datées dans les CEMA), *granica* (1 384), *horreum* (622), *granarium* (317), *canaba* (182), *spicarium* (64) et *granea* (31) sont en connexion sémantique forte avec les céréales (fig. 2). Plus d'un quart des occurrences de *granarium* est par exemple employé dans des séquences évoquant la présence de grains, en particulier *frumentum* et *bladum*, à travers les [151] redevances, au cours des XIIe et XIIIe siècles[26]. Toutefois, le contexte autour de *grangia* et de *canaba* semble plus variable que celui autour de *granarium*, *spicarium*, *granea* ou encore *horreum*, car les mentions de grains sont moins fréquentes autour de ces deux lemmes. Parallèlement, on constate que *cellarium* est beaucoup plus généraliste, avec seulement 20 associations directes avec les céréales, sur un total initial de 1 362 mentions datées[27]. Ces observations ne signifient certes pas que des grains n'étaient

---

[25] La liste des termes relatifs aux céréales a été établie à partir des dictionnaires et de l'historiographie, en particulier les travaux de G. Comet, J.-P. Devroey et M.-P. Ruas déjà mentionnés.

[26] « *et omni anno usque ad festivitatem sancti Michaelis duodecim sextarios mundissimi frumenti in granario monasterii redderent* », à Sauxillanges vers 1131 (H. Doniol (éd.), *Cartulaire de Sauxillanges*, Clermont-Ferrand, 1864, n° 4023) ; « *sextarium unum ordei in granario de Fonte Calciata annuatim in festivitate s. Felicis* », à Sant Cugat del Vallés en 1165 (J. Rius Serra (éd.), *Cartulario de Sant Cugat del Vallés*, Barcelona, 1945, n° 1052) ; ou encore « *tres modios frumenti de granario nostro* », dans un acte de Saint-Pierre de Gand en 1190 (A. Van Lokeren (éd.). *Chartes et documents de l'abbaye de Saint-Pierre au Mont Blandin à Gand depuis sa fondation jusqu'à sa suppression*, Gand, 1868-1871, n° 364 – *Diplomata Belgica* n° 801).

[27] Ces associations proviennent d'ailleurs largement du nord de l'actuelle France (Vauluisant, Laon, Ribemont, Arras, Amiens, etc.) et de la Belgique. Quelques exemples : « *in cellario fratrum daturum*

pas réservés dans des lieux désignés par le terme *cellarium* qui renvoie le plus souvent aux celliers monastiques, mais simplement que le sens de ce terme (et d'autres précédemment évoqués) allait au-delà de la question des céréales au sens strict. Présent uniquement dans les chartes italiennes, *organeum* ne semble désigner que des lieux où l'on stocke du vin[28].

Il ressort de cette analyse succincte que les lieux de stockage sont massivement présents au travers de quatre lemmes seulement, soit *granarium*, *horreum*, *granica* et *grangia*, avec les nuances évoquées. *Spicarium*, *granea*, *granatarius* évoquent sans aucun doute des réserves de grains, mais ces termes sont nettement moins fréquents dans les chartes.

[152] *Greniers terrestres et greniers célestes*

Une fois cette liste établie, il est possible d'analyser plus précisément les contextes où apparaissent les espaces de stockage. Une première lecture, qualitative, permet de relever des usages très variés, au moins jusqu'au XII[e] siècle. Dans les chartes, les réserves sont en effet mentionnées en tant qu'infrastructures[29], mais sont aussi présentes dans des énumérations stéréotypées[30], des descriptions « environnementales » où la réserve constitue un repère spatial[31], dans des clauses spécifiant le fonctionnement des redevances[32] ou encore dans une série de préambules[33].

---

*LX modios frumenti* », à l'abbaye de Stavelot, en 1126 (J. Halkin, C.G. Roland (éd.), *Recueil des chartes de l'abbaye de Stavelot-Malmédy*, Bruxelles, 1909-1930, vol. 1, n° 145 – *Diplomata Belgica* n° 1453) ; « *frumentum ad cellarium Sancte Marie cum propria uectura annuatim perducerent* », dans un acte de Barthélemy, évêque de Laon, en 1129 (A. Dufour-Malbezin (éd.), *Actes des évêques de Laon des origines à 1151*, Paris, 2001, n° 122) ; « *singulis annis in festivitate Sancti Remigii duos sextarios frumenti duos que tremesii Senonis in cellario nostro persolvetis* », acte original pour Vauluisant, en 1155 (M. Quantin (éd.), *Cartulaire général de l'Yonne, Recueil de documents authentiques pour servir à l'histoire des pays qui forment ce département*, Auxerre, 1854-1860, t. 1, n° 370).

[28] La totalité des 476 occurrences provient en effet d'Italie du Sud, en particulier de la Sainte-Trinité de Cava. Par exemple : « *debeamus vindemiare et cum vindemiatum fuerit si dederit nobis organeum debeamus* », dans une charte du chapitre cathédral de Bénévent (A. Ciaralli, V. De Donato, V. Matera (éd.), *Le più antiche carte del Capitolo della Cattedrale di Benevento (668-1200)*, Rome, 2002, n° 25).

[29] « *Actum ante granerium dicti prioratus de Chamunn* », dans un acte du prieuré de Chamonix, en 1298 (P. Lullin, C. Le Fort (éd.), « Chartes inédites relatives à l'histoire de la ville et du diocèse de Genève et antérieures à l'année 1312 », *Mémoires et documents de la Société d'histoire et d'archéologie de Genève*, t. XIV, 1862, n° 214).

[30] « *cum curte, orto, granario, vel omnis fabricis, cum suis edificiis* », dans un diplôme de Didier de Lombardie, en 757 (L. Schiaparelli (éd.), *Codice Diplomatico Longobardo*, vol. 1 et 2, Rome, Istituto storico italiano, 1929-1933, n° 127).

[31] « *id est duos campos sub horreo monachorum* », à Saint-Martin de Pontoise en 1119 (J. Depoin, J.-E. Blottière (éd.), *Cartulaire de l'abbaye de Saint-Martin de Pontoise*, Pontoise, 1895-1909, n° 24).

[32] « *horreum ad colligendas videlicet communes decimas* », dans la charte de fondation de la Trinité de Beaumont-le-Roger, en 1142 (É. Deville (éd.), *Cartulaire de l'église de la Sainte-Trinité de Beaumont-le-Roger*, Paris, 1912, n° 1).

[33] Sur les occurrences de la Genèse et l'épisode des greniers de Joseph, nous renvoyons au rapport de Michel Lauwers dans le présent volume.

Cette dernière fait en effet référence aux réserves de grain, en particulier à travers le lemme *horreum*. La surreprésentation du terme dans les textes théologiques et exégétiques a rapidement été évoquée en amont, par exemple face à *grangia* ou *granarium*. L'origine de ce décalage est à chercher dans le vocabulaire biblique, puisque les greniers y sont presque toujours désignés par le terme *horreum*, qui apparaît 20 fois dans la Vulgate, avec respectivement 16 et 6 occurrences dans l'Ancien et le Nouveau Testament[34]. À l'inverse, *grangia* et *granarium* sont absents du corpus biblique. *Horreum* [153] appartient donc à un vocabulaire plus « recherché », tout le moins sémantiquement plus chargé, que d'autres lemmes désignant des réserves de grains. Ses principaux cooccurrents dans la Vulgate sont *congrego* (6 cooccurrents) et *triticum* (4), mais encore *comburo* (3), *palleus* (2), *frux* (2), *olivum* (2) et *vinum* (2)[35]. La sémantique principale semble donc être l'accumulation des fruits de la terre (*frux*, *congrego*), en particulier du grain (*triticum*), dans un lieu spécifique (*horreum*). L'idée d'une juste redistribution est par ailleurs présente dans les versets bibliques, comme le montre Michel Lauwers dans le présent volume. À l'inverse, il ne semble jamais être question de spéculation ou de monnaie dans ces passages. Des auteurs fondamentaux, comme Augustin, Jérôme, Isidore de Séville, Bède le Vénérable ou encore Raban Maur, reprennent par la suite ces différents passages, développant une véritable théologie du stockage[36]. Le thème du grenier céleste se voit ainsi bien installé dans la pensée médiévale.

Les préambules évoquant des « réserves spirituelles » apparaissent plus tardivement dans les chartes, au XIe siècle, puis se développent principalement aux XIIe et XIIIe siècles. Reprenant différents thèmes évoqués dans la Bible, ils rappellent l'efficacité du don qui garantit à la fois un grenier terrestre abondant et une réserve céleste. Ces passages font en particulier référence aux verset des Proverbes 3 :9-10 : « *Honora Dominum de tua substantia, et de primitiis omnium frugum tuarum da ei et implebuntur horrea tua saturitate, et vino torcularia tua redundabunt* »[37]. Ils apparaissent pour la première fois à notre connaissance à

---

[34] Soit pour l'Ancien Testament quatre occurrences dans la Genèse (Gn 41:35, 41:47, 41:56, 47:22), deux dans le Deutéronome (Dt 28:5, 28:17), une dans le Livre de Ruth (Rt 2:23), deux dans les Chroniques (Ch 11:11, 31:11), deux dans le livre de Néhémie (Neh 13:12, 13:13), une dans les Proverbes (Pr 3:9-10), une dans le Livre de Joël (1:17) et enfin une dans le Livre de Malachie (Ma 3:10). Pour le Nouveau Testament, on trouve trois occurrences dans l'Évangile de Matthieu (Mt 3:12, 6:26, 13:30), et trois autres dans celui de Luc (Lc 3:17, 12:18, 12:24). Si l'on tient compte de la taille des deux ensembles (respectivement environ 580 000 et 150 000 mots), le poids des occurrences d'*horreum* est sensiblement équivalent dans les deux corpus. Les versions françaises de la Bible, comme souvent, éclatent le sens d'*horreum*, en proposant par exemple les termes « grenier » et « réserve » comme traduction – mais aussi en omettant parfois totalement de le traduire.

[35] Dans une fourchette de plus ou moins cinq mots, et avec le calcul d'un indice (Dice) permettant de faire ressortir les termes spécifiques.

[36] Par exemple : « *Audiamus Dominum per prophetam monentem: Congregate triticum in horreum meum, et sit cibus in domo mea* » (Ma 3:10), chez Agobard de Lyon, *De privilegio et jure sacerdotii*, PL 104, col. 142a.

[37] Ce même verset est repris plus de 60 fois dans la *Patrologie Latine*, tant par les Pères (Augustin, Jérôme, Isidore de Séville…), que par les auteurs des Xe-XIIe siècles (Bruno de Segni, Rupert de

Cluny, dans un acte de donation d'une vigne et de champs à l'abbaye, au début du XIe siècle[38]. L'usage du verset des Proverbes est certes plus ancien dans les CEMA, mais les occurrences antérieures ne contiennent pas la partie concernant le grenier (*horreum*)[39]. Son apparition doit donc être considérée comme [154] d'autant plus significative[40]. En dehors de Cluny, on retrouve différentes mentions du verset intégrant la référence à la réserve, par exemple à Lérins (XIe siècle)[41], à Lézat (1075-1081)[42], à Saint-Cyprien de Poitiers (1087), où il existe une variante intéressante, commandant de donner la dîme afin d'accéder au grenier céleste[43].

La majorité des préambules évoquant le stockage renvoie donc à cette idée d'un *horreum celestis*[44], au grenier du Seigneur (*horreum Domini*)[45], au grenier

---

Deutz, Pierre Damien…), en passant par ceux des époques mérovingiennes et carolingiennes (Grégoire le Grand, Bède, Raban Maur…).

[38] Durannus et sa mère Æva donnent à l'abbaye de Cluny une vigne, ainsi que des champs situés dans la *villa Canivas* : « *Quidam etiam sapiens inter cetera sua dicta ita affatur inquiens: 'Honora Dominum de tua substantia, ut ab eo benedictionem accipiant tua horrea* », dans A. Bernard, A. Bruel (éd.), *Recueil des chartes de l'abbaye de Cluny*, t. 3, Paris, 1884, n° 2064.

[39] La plus ancienne mention (874) est toutefois douteuse : « *in futuro regnum pro terrenis nobis recompensari absque ulla ambiguitate non ignoramus celestia sicut scriptum est honora dominum de tua substantia et de primitiis frugum tuarum et iterum elemosina a morte liberat et non patitur ire in infernum* », dans un pseudo-acte de Louis II le Germanique pour l'abbaye de Stavelot (P. Kehr (éd.), *Die Urkunden Ludwigs des Deutschen, Karlmanns und Ludwigs des Jüngeren*, Berlin, 1956, n° 154 (*MGH, Diplomata : Die Urkunden der deutschen Karolinger*, I)). L'occurrence fiable la plus ancienne reste à notre connaissance celle de Lézat, en 944 : « *Date helemosinam et ecce omnia munda sunt vobis et honora Dominum de tua substancia* » (P. Ourliac, A.-M. Magnou (éd.), *Cartulaire de l'abbaye de Lézat*, Paris, 1984, n° 1673). On a, par la suite, pu relever plus de 40 mentions contenant le syntagme « *Honora dominum de tua substantia* » dans les CEMA, à Lézat, Lérins, Cluny, Baigne, Cambrai, La Charité-sur-Loire, Sylvanès, Valmagne, etc., entre les années 990 et 1230. À Cluny aussi, des mentions du verset précèdent celle contenant la partie sur *horreum* : A. Bernard, A. Bruel (éd.), *Recueil des chartes de l'abbaye de Cluny…*, n° 324 (927-942) et 1143 (janvier 963).

[40] L'absence pourrait s'expliquer par une copie incomplète des préambules lors de la rédaction des cartulaires. Cette hypothèse nous paraît toutefois plutôt faible, d'une part parce que d'autres passages bibliques accompagnent le verset des Proverbes dans les chartes antérieures et qu'ils n'ont pas été coupés, d'autre part, parce que l'hypothèse d'une troncature plus ou moins systématique des préambules au moment de la cartularisation n'a jamais été démontrée, et encore moins à l'échelle européenne.

[41] « *item per Salomonem* : « *honora Dominum de tua substantia et de primiciis frugum tuarum ut impleantur orrea vestra saturitate et vino torcularia redundabunt* », dans H. Moris, E. Blanc (éd.), *Cartulaire de l'abbaye de Lérins*, Paris, t. 1, 1883, n° 53bis (*De Barjamone*).

[42] « *In veteri et in novo Testamento precepit omnipotens Deus ut unusquisque homo faciat de omnibus suis rebus helemosinas et honoret Dominum de omni sua substancia ut inpleantur in presenti seculo orrea eorum saturitate et in futuro possideant perpetuam felicitatem* », dans P. Ourliac, A.-M. Magnou (éd.), *Cartulaire de l'abbaye de Lézat...*, n° 1062.

[43] « *Lex divina precipit honorare Dominum Deum dicens honora Dominum Deum tuum de tua substantia et de decimis frugum tuarum da pauperibus ut impleantur orrea tua in celis* », dans L. Rédet (éd.), *Cartulaire de l'abbaye de Saint-Cyprien de Poitiers*, Poitiers, 1874, n° 598.

[44] « *Quia quicumque thesauris mundane conversationis delectatur, eterne securitatis tranquillitate priuatur. His ammonicionibus aliquantulum excitatus, eterne epulationis aliqua grana, in orreum celeste, in eternum uicturum vellem* », dans un acte de Godefroi de Conversano en 1093 (A. Spinelli (éd.), *Monumenta regii Neapolitani archivi edita ac illustrata*, vol. 5, Naples, 1857, n° 470).

[45] Voir les documents mentionnés précédemment.

éternel (*horreum eternitatis*)[46], allant parfois jusqu'à comparer l'institution ecclésiale à un grenier, tout en faisant référence à la récolte [155] des âmes, comparées à des grains – comme dans un acte pontifical de Clément V pour l'université de Montpellier (1309)[47]. Ces occurrences se rencontrent essentiellement aux XIIe et XIIIe siècles, plutôt dans des bulles, mais aussi dans des textes émanant de grands personnages, à l'abbaye de Tronchiennes[48], chez les évêques de Metz[49], ou encore à Saint-Gall[50]. Relativement nombreuses, ces mentions sont intéressantes car elles disent quelque chose de l'évolution de la pensée autour de la mise en réserve au cours des Xe-XIIIe siècles. Le grenier se voit alors intégré, dans les chartes (car la chose est plus classique en exégèse), à la rhétorique ecclésiale. L'essentiel de la documentation est cependant ailleurs, dans les structures de stockage elles-mêmes et leurs environnements, leurs usages pratiques et les formules énumératives. Comment s'articulent la chronologie des greniers célestes et celle des greniers terrestres ? Surtout, comment évaluer la répartition chronologique de ces dispositifs ?

**Chronologie relative des structures de stockage**

*Méthode et première lecture*

Les corpus historiques sont évidemment répartis inégalement d'une période à l'autre, ce qui complique les analyses diachroniques. Si l'on s'en tient aux données brutes, il y a donc un risque de mesurer l'évolution du corpus lui-même et non celle de la fréquence des mots. Une solution statistiquement satisfaisante pour contourner cette difficulté consiste à diviser en « tranches » chronologiques le corpus, de façon à ce que chacune [156] d'entre elles contienne une quantité de mots équivalente[51]. On forme ainsi des « tranches », correspondant à des périodes plus ou moins longues : il faut parfois un siècle pour réunir 1 million de mots consécutifs (en particulier pour le haut Moyen Âge), parfois seulement dix années. Une fois le corpus divisé, il s'agit de compter le nombre d'occurrences des termes recherchés pour chacune des tranches, afin de mesurer l'évolution des mentions, indépendamment du corpus lui-même.

---

[46] « *Ne flos humane laudis velocius umbra pertransiens juventutem nostram ex toto teneret captivam qui non dormit neque dormitat pium loquor Ihesum ut fructum in horreo eternitatis reponendum faceret interiorem hominem nostrum excitavit* », dans un acte pour l'abbaye de Vaucelles en 1166 (B.-M. Tock (éd.), *Les chartes de l'abbaye cistercienne de Vaucelles au XIIe siècle*, Turnhout, 2010, n° 40).

[47] « *in Ecclesie firmamento multorum diversitatem fidelium, qui divitias scientiarum amabiles in sinu ejusdem Ecclesie congregent spatioso, et in horreum Domini grana salutis inferant* » (1309), dans É Baluze, G. Mollat (éd.), *Vitae paparum Avenionensium. Nouvelle édition revue d'après les manuscrits et complétée de notes critiques*, Paris, 1914-1922, vol. 3, n° 26.

[48] « *inter omnia hujus mundi bona ea solum in vere beatitudinis horreum reservari ac perpetuari que vel in ecclesiarum edificationem vel pauperum christi sustentationem liberaliter erogantur* », dans un acte de Thierry II d'Alsace pour l'abbaye (T. De Hemptinne, A. Verhulst (éd.), *De oorkonden der graven van Vlaanderen (juli 1128-september 1191) II. Uitgave. Band I: regering van Diederik van de Elzas (juli 1128-19 januari 1168)*, Bruxelles, 1988, n° 45).

[49] « *Cum his vero supra nominatis ne ad horreum domini omnino vacui accedamus pro remedio anime nostre* », dans un acte original (multi-édité) de l'évêque de Metz, Étienne, pour l'abbaye de Saint-Benoît-en-Woëvre (M. Parisse (éd.), *Les actes des évêques de Metz*, Metz, 1964, n° 41).

[50] « *Homo libere conditionis usus consilio summi pontificis vestri antecessoris pro salute mea meorum que statui aliqua de meis mittere in horreum domini* », dans un acte original de Diethelm de Toggenburg pour le pape Innocent III, en 1201 (O. P. Clavadetscher (éd.), *Chartularium Sangallense. Band III, 1000-1265*, Saint-Gall, 1983, n° 971).

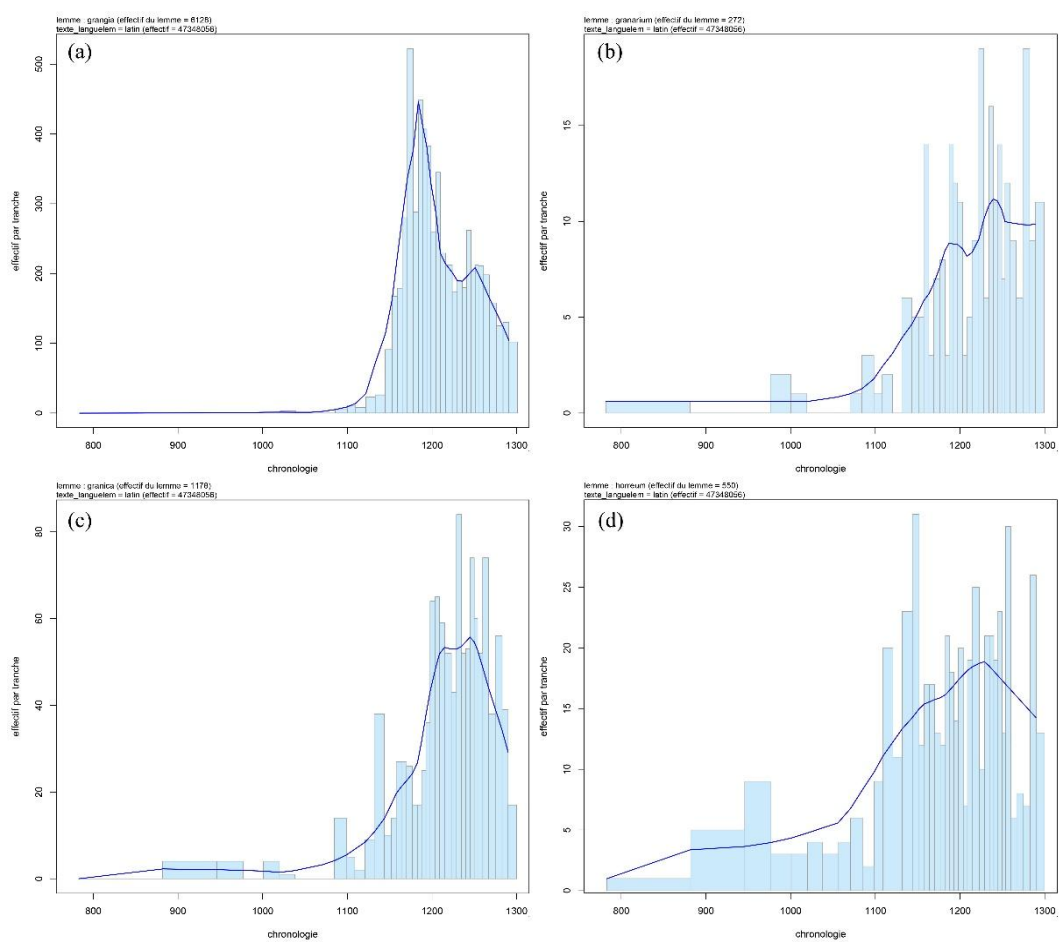

---

[51] Cette méthode a déjà été employée dans N. Perreaux, « Chronologie, diffusion et environnement des *villae*… » ; *id.*, « Des structures domaniales aux territoires ecclésiaux ? Entités spatiales et dynamique du processus de spatialisation dans les actes diplomatiques (VII<sup>e</sup>-XIII<sup>e</sup> siècles) », T. Martine, J. Nowak, J. Schneider (dir.), *Espaces ecclésiastiques et seigneuries laïques. Définitions, modèles et conflits en zones d'interface (IX<sup>e</sup>-XIII<sup>e</sup> siècle)*, Paris, à paraître. Une fonction de la bibliothèque Cooc, développée par Alain Guerreau pour le logiciel d'analyses statistiques R, permet d'automatiser ce type de recherche sur un corpus au format CQP.

**Fig. 3.** Chronologie des mentions de (a) *grangia*, (b) *granarium*, (c) *granica* et (d) *horreum* dans les CEMA, 800-1300.

[157] On constate ainsi que l'évolution des quatre principaux lemmes désignant des lieux associés aux grains – *grangia*, *granarium*, *granica* et *horreum* – est particulièrement nette et concordante (fig. 3). Les occurrences, rares, voire très rares dans le haut Moyen Âge, augmentent lentement, d'abord aux X$^e$ et XI$^e$ siècles pour *horreum*, puis explosent littéralement aux XII$^e$-XIII$^e$ siècles. On pourrait arguer que cette situation était attendue pour *grangia*, à travers le développement des Cisterciens[52]. C'est toutefois sans compter avec le fait que les premières mentions du lemme dans les chartes apparaissent chez les Bénédictins, et plus précisément à Cluny, dans la décennie 970 – puis à Savigny, Saint-Aubin d'Angers, Marmoutier, etc.[53] Enfin, au XIII$^e$ siècle, le plus souvent vers 1250, les occurrences de ces lieux chutent assez brutalement.

*Une absence des fosses-silos dans les textes ?*

Comment interpréter cette relative rareté pour le haut Moyen Âge puis cette explosion du lexique du stockage des grains aux XII$^e$-XIII$^e$ siècles ? Un tel résultat paraît *a priori* contre-intuitif : les archéologues observent une forte présence des fosses-silos (au moins) dès les VI$^e$-VIII$^e$ siècles[54]. Tout se passe donc comme si les fouilles et le corpus textuel divergeaient [158] ou plutôt se complétaient[55]. Cette différence paraît tenir d'abord à deux éléments, qu'il convient d'ailleurs d'articuler : d'une part, la typologie des structures mentionnées dans les textes, par rapport à celles découvertes en fouille ; d'autre part, la sémantique des chartes et l'évolution du rapport à la mise en réserve, dans une société où les rapports de domination se transforment[56].

---

[52] Voir de nouveau l'analyse de D. Panfili, « *Domus*, *grangia*, *honor* et les autres… », en particulier p. 320-325. Il est certes vrai qu'une bonne partie des mentions de *grangia* provient des textes cisterciens (même si, comme le montre Didier Panfili, celles-ci ne sont pas omniprésentes dans les chartes cisterciennes du Languedoc). On pourrait toutefois se demander si ces occurrences sont propres à un ordre monastique ou si elles traduisent certaines évolutions sociales de la seconde moitié du XII$^e$ siècle.

[53] « *Et in alio loco unam perticam que habet in longum perticas III inante grangiam jacet* », dans un don de Wichardo et Ademaro à l'abbaye de Cluny, en février 974 (A. Bernard, A. Bruel (éd.), *Recueil des chartes de l'abbaye de Cluny…*, n° 1360) ; « *unum pratum simul cum fratre meo Iarentone et est grangia ipsorum monachorum juxta ipsum pratum* » puis « *vinea mansionem ipsius Godalfredi cum grangia ipsius* », à l'abbaye de Savigny en 1020 (A. Bernard (éd.), *Cartulaire de l'abbaye de Savigny suivi du petit cartulaire de l'abbaye d'Ainay*, Paris, 1853, n°653) ; « *dedit pro hac re supradicto Rotberto grangiam suam totam integram sicut edificata erat* », à Saint-Aubin d'Angers en avril 1029 (E. Lelong, A. Bertrand De Broussillon (éd.), *Cartulaire de l'abbaye de Saint-Aubin d'Angers*, Paris, 1896-1903, n° 350).

[54] É. Peytremann, « Structures et espaces de stockage dans les villages alto-médiévaux… », p. 43-45, 47, 53 ; M. Bois, L. Schneider, « Conclusion », O. Maufras (éd.), *Habitats, nécropoles et paysages dans la moyenne et basse vallée du Rhône (VII$^e$-XV$^e$ siècle)…*, p. 427-445 ; L. Schneider, « De la fouille des villages abandonnés à l'archéologie des territoires locaux. L'étude des systèmes d'habitat du haut Moyen Age en France méridionale (V$^e$-X$^e$ siècle). Nouveaux matériaux, nouvelles interrogations », J. Chapelot (dir.), *Trente ans d'archéologie médiévale en France. Un bilan pour un avenir*, Caen, p. 133-161 ; R. Carme, Y. Henry, « L'ensilage groupé et les campagnes du premier Moyen Âge dans le Toulousain : quelques réflexions à l'aune de deux fouilles récentes (L'oustalou à Préserville et Clos-Montplaisir à Vieille-Toulouse) », *Archéologie du Midi médiéval*, vol. 28, 2010, p. 33-101.

L'analyse des CEMA n'a, en effet, pas permis de repérer de mention faisant explicitement référence aux pratiques d'ensilage souterrain, qui semblent *de facto* quasi-invisibles dans les chartes[57]. La fameuse occurrence du capitulaire *De Villis* (vers 800), interdisant la dissimulation de grains, soit en silos enterrés (*subtus terram*), soit ailleurs (*aliubi*), nous semble à verser à ce dossier car elle montre l'existence de pratiques d'ensilage individuelles ou collectives peu régulées[58]. Parallèlement, de nombreux éléments tendent à montrer qu'*horreum, granarium, granea* ou encore *grangia* [159] renvoient systématiquement à des constructions pérennes, en bois, voire en pierre. Ainsi, dans l'actuelle France du nord et en Belgique, mais aussi en Italie et en Bourgogne, on relève différentes mentions prévoyant l'approvisionnement en bois pour la construction de ces structures[59], leur réparation[60], par exemple en cas de destruction par le feu[61], mais encore l'intervention d'un charpentier, comme dans cet acte de Gorze en 984 : « *Si domus indominicata aut horreum destructum fuerit, cum carpentario nostro ipsi restaurabunt* »[62]. Ces deux informations combinées laissent penser que seules les structures pérennes, construites en bois ou en pierre, sont mentionnées dans les chartes. Ce qui explique en définitive leur rareté, voire leur absence aux VII[e]-X[e] siècles.

---

[55] Le phénomène avait déjà été relevé de façon non systématique par Édith Peytremann, dans « Structures et espaces de stockage dans les villages alto-médiévaux… », p. 42 : « Il convient au demeurant de souligner que, pour cette période [les VIe-VIIe siècles], aucune allusion au stockage en silos n'est mentionnée dans les lois, du moins sous les formes employées par les agronomes latins tels Pline, Columelle et Varron » ; puis p. 47 : « [L]'ensilage ressort comme le mode de stockage le plus utilisé durant le premier Moyen Âge. Ce résultat contraste grandement avec les sources écrites et incite à la prudence ».

[56] J. Demade, *Ponction féodale et société rurale en Allemagne du sud (XIe-XVIe siècles). Essai sur la fonction des transactions monétaires dans les économies non capitalistes*, Strasbourg, 2004.

[57] Avec une possible exception : pour la Catalogne, Roland Viader nous signale que certains termes (les formes *ciga, cigia, cia, ecia, scia, citgia, sitgia*, mais aussi le lemme *fovea* – probablement dérivé de *fossa*) pourraient renvoyer à des silos. Ces occurrences sont majoritairement présentes aux Xe et XIe siècles. On trouve en effet quelques dizaines de formes liées aux *sitjes* dans les CEMA. Le lemme *fovea* est pour sa part plus présent, mais il possède des significations très variables, renvoyant à toutes sortes de structures en creux (bassefosses, pêcheries, lieu de stockage, etc.). Voir V. Farías, R. Martí, A. Catafau, *Les sagreres a la Catalunya medieval*, Girona, 2007, p. 34 ; A. Catafau, « Les *celleres* du Roussillon. Le regroupement villageois dans l'espace consacré autour de l'église et son rôle dans la formation de l'habitat concentré dans l'ancien diocèse d'Elne, Xe-XIVe siècles », P. Sénac (dir.), *Histoire et archéologie des terres catalanes au Moyen Âge*, Perpignan, 1995, p. 163-195.

[58] « *Praevideat unusquisque judex, ut sementia nostra nullatenus pravi homines subtus terram vel aliubi abscondere possint et propter hoc messis rarior fiat* », dans A. Boretius, (éd.), *Capitularia regum Francorum*, t. 1, Hannovre, 1883, n° 32:51, p. 88. Sur le capitulaire et sa datation, voir F.-L. Ganshof, « Observations sur la localisation du *Capitulare de villis* », *Le Moyen Âge*, vol. 5, 1949, p. 201-223 ; J.-P. Devroey, *Économie rurale et société dans l'Europe franque (VIe-IXe siècles)*, Paris, 2003, p. 197. Une nouvelle édition des capitulaires est actuellement en cours, sous la direction de Karl Ubl (<https://capitularia.uni-koeln.de/>), et un volume collectif doit paraître prochainement sur la question : B. Jussen, K. Ubl (dir.), *Die Sprache des Rechts. Historische Semantik und karolingische Kapitularien*, Göttingen, 2019.

[59] Par exemple : « *Concessi quoque ut ad omnia edificia prefate grangie ligna accipiendi liberam* », à Cîteaux en 1178 (J. Marilier (éd.), *Chartes et documents concernant l'abbaye de Cîteaux, 1098-1182*, Rome, 1961, n° 232) ; « *Si domum vel horreum super ipsum allodium edificare voluerit* (…) *in silva sancti Aychadri habebit ligna* », à Ninove en 1184 (J. J. De Smet (éd.), *Recueil des chroniques de Flandre*, Bruxelles, 1837-1841, n° 48 ; *Diplomata Belgica*, n° 4784).

*Les occurrences du haut Moyen Âge : chartes et inventaires*

Les chartes du haut Moyen Âge mentionnent certes un certain nombre de lieux de stockage de grains. Mais si l'on passe à une lecture qualitative des textes, on constate qu'une majorité de ces rares mentions antérieures à 850 provient de pseudo-inventaires ou de listes de biens. On relève par exemple trois occurrences de *granarium* dans le procès-verbal de l'insinuation des *Gesta municipalia* de Ravenne, en 564 – où le terme désigne [160] d'ailleurs des coffres à grains[63] – mais encore un *horreum* dans un acte très énumératif de l'évêque de Chartres Ageradus, pour l'abbaye Notre-Dame, en 696[64]. À l'inverse, il est remarquable qu'aucune mention de *granarium* ou d'*horreum* n'apparaisse dans les diplômes mérovingiens, authentiques comme faux[65]. Au VIII[e] siècle et dans la première moitié du IX[e] siècle, en dehors des pseudo-inventaires ou des chartes interpolées, comme le testament de Chrodegang pour l'abbaye de Gorze[66], seuls deux documents lombards mentionnent des dispositifs pour la réservation des grains, à travers des formules énumératives (757, 764)[67]. Il en va de même dans la première moitié du IXe siècle, où les quelques occurrences proviennent, en fait, toutes de faux[68]. C'est donc après 850 que les occurrences deviennent moins sporadiques, par exemple à l'abbaye de La Grasse en 882, avec un document conservé en original[69], mais aussi à Autun, Nevers, Lézat, Conques, Auch, Saint-Bénigne de Dijon, Langres, etc.[70] Tout se passe donc comme si les chartes typiques des VIIe-IXe siècles, [161] du moins celles renfermant des dons et des confirmations, n'évoquaient pas ces lieux de réserve (fig. 4).

---

[60] « *Nullam struem de lignorum collectione (...) ad reparandam propriam domum, proprium horreum* », à Saint-Pierre de Gand en 1207 (A. Van Lokeren (éd.), *Chartes et documents de l'abbaye de Saint-Pierre au Mont Blandin à Gand…*, n° 424).

[61] « *pro restauratione grangie vestre combuste* », dans un privilège d'Urbain III pour l'abbaye de Mortemer, en 1186 (P.F. Gallagher (éd.), *Monastery of Mortemer-en-Lyons in the Twelfth Century: its History and its Cartulary*, Indiana, 1970, n° 12) ; « *Armata enim manu atrium uiolauit sanguinem in loco sacro fudit parietes templi ligonibus infregit et suffodit horrea nostra et domos canonicorum nostrorum incendit* », dans un jugement du comte de Flandres en faveur de l'évêque de Thérouanne (T. De Hemptinne, A. Verhulst (éd.), *De oorkonden der graven van Vlaanderen…*, n° 64). L'ensemble des occurrences semble montrer que l'incendie d'une grange a souvent servi de représailles en cas d'aggravation de conflit seigneurial.

[62] A. d'Herbomez (éd.), *Cartulaire de l'abbaye de Gorze Ms. 826 de la Bibliothèque de Metz*, Paris, 1898-1900, n° 116. Il s'agit d'une confirmation des privilèges des habitants de Brouck par l'abbé Ermenfridus.

[63] « *arca granaria minore ferro legata valente siliquas aureas duas* », dans *Chartae Latinae Antiquiores*, Zurich, 1954-, t. 17, n° 652 (fac-similé p. 10-12, 14-18, 20-22) ; Artem, n° 1762 ; papyrus original conservé dans Paris, BnF, ms. Latin 4568 A. Sur ce type documentaire, voir J. Barbier, *Archives oubliées du haut Moyen Âge. Les* gesta municipalia *en Gaule franque (VI[e]-IX[e] siècle)*, Paris, 2014.

[64] « *triticum horrea reconditis metere atque adepisci mancipari valeamur* », dans *Chartae Latinae Antiquiores…*, t. 14, n° 580 ; Artem n° 4475.

[65] C. Brühl, T. Kölzer (éd.), *Die Urkunden der Merowinger*, Hanovre 2001.

[66] « *semina de curte accipere, metere et in horreum ducere, atque triturare, horreum cooperire* », dans le pseudo-testament daté du 25 mai 765 (cf. A. d'Herbomez (éd.), *Cartulaire de l'abbaye de Gorze…*, n° 11). Pour la critique du document, voir P. Marichal, *Remarques chronologiques et topographiques sur le Cartulaire de Gorze*, Paris, Klincksieck, 1902, p. 18-19.

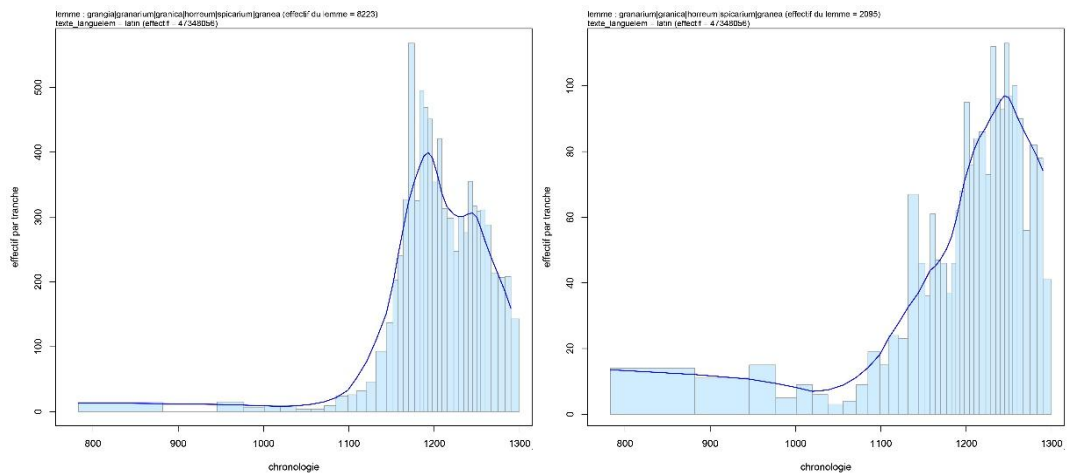

**Fig. 4.** Chronologie des mentions cumulées de (a) *grangia* + *granarium* + *granica* + *horreum* + *spicarium* + *granea* dans les CEMA, 800-1300 ; (b) *idem*, mais sans *grangia*.

Ce contraste entre les chartes du haut Moyen Âge et les découvertes archéologiques nous semble imputable à la nature même des rapports sociaux décrits dans ces textes. Pour cette période, et contrairement à ce que l'on trouve par exemple dans les inventaires (donc les polyptyques), les dominants ne cherchent généralement pas à y rapporter comment on produit (voire même ce que l'on produit) mais s'intéressent quasi-exclusivement au don et aux conséquences sociologiques de cette pratique[71]. Cela n'exclut pas la rationalité de ces acteurs, mais indique simplement que cette rationalité semble alors s'exercer sur d'autres champs sociaux que le stockage, en particulier que le stockage en silos[72]. L'examen du [162] corpus des capitulaires dans son ensemble confirme cette impression, faisant apparaître qu'une majorité des rares occurrences de réserve de grains provient du *Brevium exempla ad describendas res ecclesiasticas*, daté d'environ 810. Celui-ci est d'ailleurs bien plus un modèle d'inventaire qu'un capitulaire normatif[73]. Autrement dit, il semble non seulement que les lieux de stockage aient été plus rares dans le haut Moyen Âge qu'au Moyen Âge central (une observation *a priori* attendue), mais aussi que ces réserves, parfois enterrées, n'étaient pas objectivées par les discours dominants. Donc qu'elles n'étaient probablement que rarement la cible directe des ambitions seigneuriales. On peut alors faire l'hypothèse que les puissants ne contrôlaient généralement pas le stockage, sauf dans certains grands centres, au moins jusqu'au X[e] siècle.

---

[67] Par exemple, « *excepta vehicione vini et frumenti undecumque in usus fratrum eorum que portacione in granarium atque cellarium* », dans un pseudo-acte de 812, attribué à Charlemagne, pour Neustadt (E. Mühlbacher, A. Dopsch, J. Lechner, M. Tangl (éd.), *Die Urkunden Pippins, Karlmanns und Karls des Großen*, Hanovre, 1906, n° 283).

[68] « *cum curte, orto, granario, vel omnis fabricis* », dans un diplôme de Didier de Lombardie, daté du 5 novembre 757 (L. Schiaparelli (éd.), *Codice Diplomatico Longobardo…*, n° 127) ; « *cum curte et puteum, cum granario et ipsa sala* », dans un diplôme de 764, par le même (*ibid.*, n° 178).

[69] « *ipsa porta Redensis solarios II, ubi nos abitamus, cum ipso orreo et cum illorum curte.* », É. Magnou-Nortier, A.-M. Magnou (éd.), *Recueil des chartes de l'abbaye de la Grasse*, Paris, 1996-2000, n° 32 ; Artem n° 3780.

[70] Quelques exemples : « *omnis decimatio frugum, que ad orreum et cellarium dominicum veniunt* », dans un acte de Charles III le Gros pour Saint-Martin de Nevers, en 886 (P. Kehr (éd.), *Die Urkunden Karls III*, Hanovre, 1936-1937, n° 138) ; « *seticum indominicatum supra fluvium Ararim cum granea et horto et curti* », à Saint-Nazaire d'Autun, en 910 (A. De Charmasse (éd.), *Cartulaire de l'Église d'Autun*, Paris, 1865, n° 34) ; « *cum omnes apenditiones eorum, cum casas et orrea et terras vel vineas* », à Lézat en 948 (P. Ourliac, A.-M. Magnou (éd.), *Cartulaire de l'abbaye de Lézat…*, n° 609). Cette évolution est d'ailleurs visible sur la fig. 3, où l'on note par exemple un développement d'*horreum* vers 870-880.

## Un basculement socio-sémantique majeur (X[e]-XIII[e] siècle)

### *L'essor des mentions de grains*

L'explosion des structures de stockage pérennes aux XII[e]-XIII[e] siècles paraît donc relever d'une toute autre logique que celle du haut Moyen Âge. C'est cette hypothèse de travail que la suite de l'exposé propose d'explorer, à partir d'enquêtes sémantiques. En premier lieu, il convient de regarder de plus près l'évolution des mentions de grains. Une liste de lemmes les plus fréquents a été composée à partir des dictionnaires et de l'historiographie[74] : *siligo / sigale* (pour le seigle, soit 2 296 mentions), [163] *hordeum* (pour l'orge, 2 670), *triticum-cibaria* (pour le froment, 1 768), *spelta* (pour l'épeautre, 417), *avena / civata* (pour l'avoine, 7 764), *frumentum* (pour le froment, mais aussi pour les grains en général, 9 460), *milium / panicum* (pour le millet, 354) et enfin *bladum* (qui désigne aussi différents grains, 5 939)[75] ; soit plus de 30 500 mentions sur l'ensemble des CEMA, tout en sachant qu'il existe des incertitudes et des recouvrements quant aux céréales désignées par ces termes.

---

[71] Sur les liens entre « don » et « prélèvements », voir L. Kuchenbuch, « *Porcus donativus*. Sprachgebrauch und Gabenpraxis in der seigneurialen Überlieferung vom 8. zum 12. Jahrhundert », G. Algazi, B. Jussen, V. Groebner (dir.), *Negotiating the Gift: Pre-Modern Figurations of Exchange*, Göttingen, 2003, p. 203-256 ; E. Magnani, « Les médiévistes et le don. Avant et après la théorie maussienne », E. Magnani, *Don et sciences sociales. Théories et pratiques croisées*, Dijon, 2007, p. 15-28 ; J. Morsel, « Le prélèvement seigneurial est-il soluble dans les Weistümer ? Appréhensions franconniennes (1200-1400) », M. Bourin, P. Martinez Sopena (dir.), *Pour une anthropologie du prélèvement seigneurial dans les campagnes de l'Occident médiéval (XI[e]-XIV[e] siècles). Réalités et représentations paysannes*, Paris, 2004, p. 155-210 ; ainsi que la note suivante.
[72] Différents auteurs, à la suite de Marcel Mauss, de Maurice Godelier, puis d'Alain Guerreau ont montré combien ce que nous appelons aujourd'hui l'« économie » n'avait qu'une faible autonomie sociale au cours des périodes qui nous retiennent (cf. D. Iogna-Prat, « Préparer l'au-delà, gérer l'ici-bas. Les élites ecclésiastiques, la richesse et l'économie du christianisme (perspectives de travail) », R. Le Jan, L. Feller, J.-P. Devroey (dir.), *Les élites et la richesse au haut Moyen Âge*, Tunhout, 2010, p. 59-70). Consécutivement, on s'interroge donc sur la place de la « production » et des « prélèvements », qui semblent eux-mêmes largement dépendants de la question de la domination, donc des relations interpersonnelles et de la parenté. Bien souvent, il semble que la relation de *dominium* impliquait, du moins pour le haut Moyen Âge, la capacité de prélever – sans que la nature, le mode, le temps et la quantité de ce prélèvement soient fixés par écrit (du moins dans les chartes). Sur ce point, voir J. Demade, *Ponction féodale et société rurale…* Autres éléments de discussions dans V. Toneatto, « Élites et rationalité économique : les lexiques de l'administration monastique du haut Moyen Âge », R. Le Jan, L. Feller, J.-P. Devroey (dir.), *Les élites et la richesse au haut Moyen Âge…*, p. 71-96 ; J.-P. Devroey, *Puissants et misérables. Système social et monde paysan dans l'Europe des Francs (VI[e]-IX[e] siècles)*, Bruxelles, 2006, p. 588 et suiv.
[73] Sur les capitulaires et les polyptyques, voir le texte de Jean-Pierre Devroey dans le présent volume.

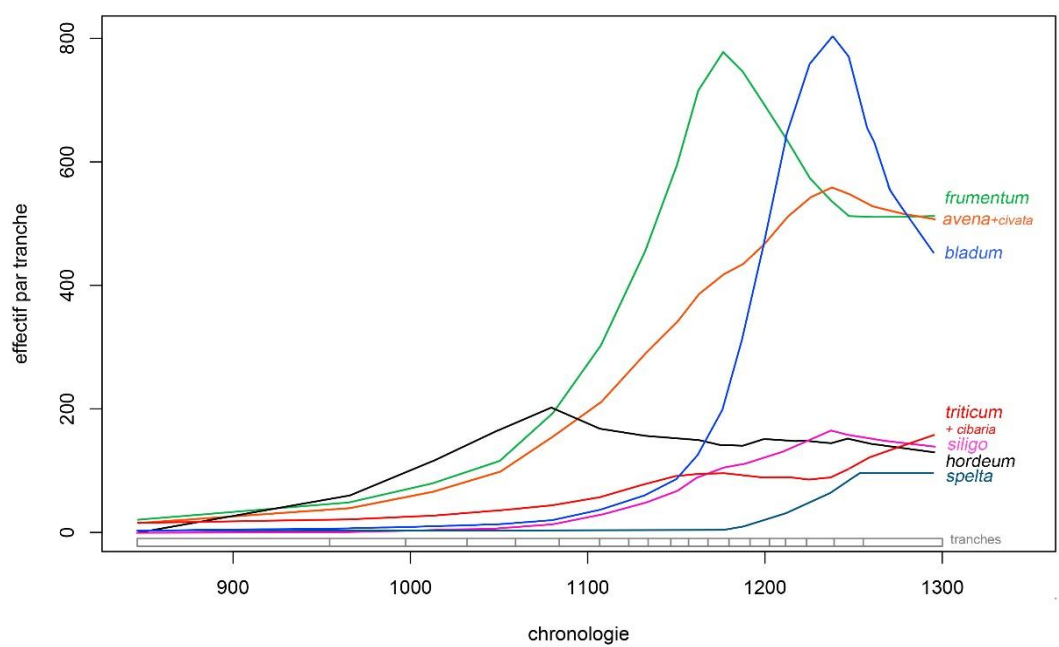

**Fig. 5.** Chronologie des occurrences de *frumentum*, *avena* / *civata*, *bladum*, *triticum* / *cibaria*, *hordeum* et *spelta* dans les CEMA (milieu VIIIe-XIIIe siècle).

Après application de la même méthode d'analyse diachronique, les différents termes ont été placés sur un graphique de synthèse (fig. 5). Celui-ci permet d'observer que l'évolution des mentions de lieux de stockage (fig. 3 et 4) est parallèle à la dynamique des occurrences de grains, [164] avec un très léger décalage chronologique pour les mentions de lieux. Les occurrences de céréales connaissent, en effet, un essor phénoménal aux XII[e]-XIII[e] siècles, dans un moment où s'affirment certaines formes du contrôle seigneurial. On constate que cet essor est aussi lié à une évolution dans la typologie des grains, du moins dans les chartes (fig. 5)[76].

---

[74] G. Comet, *Le Paysan et son outil…*, p. 215-217 ; J.-P. Devroey, « La céréaliculture au haut Moyen Âge... », p. 64 ; *id.*, « La céréaliculture dans le monde franc », *L'ambiente vegetale nell'alto Medioevo*, Spolète, 1990, p. 221-256 ; ainsi que les très riches travaux de M.-P. Ruas, dont *Productions agricoles, stockage et finage en Montagne Noire médiévale. Le grenier castral de Durfort (Tarn)*, Paris, 2002, p. 207-218.

[75] On notera l'importance très inégale des différents lemmes, entre *spelta* (417 mentions) et *frumentum* (9 640). Les occurrences de millet sont les plus rares dans le corpus, avec les lemmes *milium* (189 mentions) et *panicum* (165). Ces termes sont d'ailleurs présents quasi-exclusivement dans les chartes du sud de l'Europe, en Italie, Provence, Languedoc et Espagne. Cette régionalisation est plus forte encore pour *panicum*, présent essentiellement en Italie.

Jusqu'au milieu du XI[e] siècle, une certaine concurrence règne entre les variétés, sauf pour l'orge (*hordeum*) qui se distingue légèrement entre environ 980 et 1080[77]. S'observe par la suite une évolution à la hausse pour pratiquement tous les lemmes du champ lexical. Le XII[e] siècle voit néanmoins surtout s'affirmer *frumentum*, *avena* et *bladum*, qui désignent d'abord l'avoine et le froment – même si *bladum* et dans une moindre mesure *frumentum* sont parfois génériques[78]. De tels indicateurs doivent certes être pris avec précaution : la dynamique des occurrences observée doit beaucoup à l'intérêt seigneurial pour certains types de redevances et pour le pain blanc, « symbole de distinction sociale », comme le rappelle Jean-Pierre Devroey[79]. Malgré ces questions en suspens, deux points [165] paraissent

---

[76] La répartition des espèces observées par M.-P. Ruas au sein du grenier castral de Durfort est très différente (M.-P. Ruas, *Productions agricoles, stockage et finage en Montagne Noire médiévale…*), mais il s'agit d'un lieu particulier, dans une chronologie différente de la nôtre (XIV[e] siècle). À ce stade, il reste donc difficile de conclure à une différence entre les espèces observées et celles mentionnées dans les chartes, même si cela reste probable. Les chartes ne reflètent en effet pas le « quotidien » des populations médiévales, du moins pas prioritairement, mais la dynamique du système, donc des espaces qu'elles renseignent, ainsi que bien souvent les attentes de la frange dominante de la population. Il ne s'agit donc pas de discuter leur « représentativité » face aux observations archéologiques, mais plutôt de dire qu'elles sont le reflet de tendances sociales et de discours orientés, complémentaires des fouilles.

[77] Si les mentions d'*hordeum* sont plutôt stables par la suite, c'est toutefois le seul lemme désignant des grains qui n'augmente pas dans le corpus entre 1080 et 1300.

[78] Pour une vision globale de l'évolution de l'avoine de la Protohistoire au XV[e] siècle, à partir des fouilles, voir M.-P. Ruas, V. Zech-Matterne, « Les avoines dans les productions agro-pastorales du nord-ouest de la France. Données carpologiques et indications textuelles », V. Carpentier, C. Marcigny (dir.), *Des hommes aux champs. Pour une archéologie des espaces ruraux du Néolithique au Moyen Âge*, Rennes, 2012, p. 327-365.

[79] « La farine blanche, la plus blanche, est restée continûment une nourriture aristocratique, clairement et massivement perçue comme telle : les ascètes et les pénitents y renoncent, les citadins, même modestes, y recherchent systématiquement une expression de leur supériorité. », dans A. Guerreau, « L'étude de l'économie médiévale : genèse et problèmes actuels », J. Le Goff, G. Lobrichon (dir.), *Le Moyen Âge aujourd'hui. Trois regards contemporains sur le Moyen Âge : histoire, théologie, cinéma*, Paris, 1998, p. 31-82 ; J.-P. Devroey, « La céréaliculture au haut Moyen Âge…, p. 60 ; J.-P. Devroey, « Entre Loire et Rhin : les fluctuations du terroir de l'épeautre au Moyen Âge », J.-P. Devroey, J.-J. Van Mol (dir.), *L'épeautre, histoire et ethnologie…*, p. 89-106 ; *id.*, *Économie rurale et société…*, p. 104-105, où l'auteur observe qu'à Saint-Remi de Reims, l'épeautre occupe 91,3 % des terres céréalières, et le seigle 6,9 % (soit 98,2 % du total pour ces deux espèces). Toutefois, ces proportions sont très différentes dans d'autres inventaires, par exemple au début du X[e] siècle à Sainte-Julie de Brescia, où l'on a 39 % de seigle et 20 % de froment, sans mention d'épeautre. Le dernier cas indiqué par l'auteur est celui de Saint-Martin de Tours, avec 38 % d'orge, 27 % de froment, 18 % d'avoine et 15 % de seigle. Il ne faut pas oublier que les chartes reflètent avant tout le discours et l'idéologie d'une frange limitée de la population, dans des contextes orientés (souvent celui du don à l'Église). Parallèlement, la régionalisation du phénomène

relativement clairs, du moins à l'échelle macro[80] : 1. certaines espèces de céréales s'affirment et deviennent majoritaires dans les chartes à partir du XII[e] siècle, en particulier le froment et l'avoine ; 2. l'essor très net des mentions de grains est parallèle à celui des lieux de stockage, avec peut-être quelques décennies d'avance (dès 1050-1075 environ). Cela semble signifier que l'intérêt pour les grains et leur répartition/captation a accompagné, voire précédé, le gros de la création des structures de stockage en bois et en pierre, relevées dans les chartes.

Dans le détail, il semble possible de distinguer quatre phases chronologiques (fig. 5). La première s'étendrait du VIII[e] au début du X[e] siècle[81]. Lors de celle-ci, les chartes évoquent essentiellement le froment et l'avoine, mais les mentions d'orge augmentent aussi lentement. La quasi-absence d'épeautre et de seigle est, lors de cette première période, remarquable, du moins face aux observations archéologiques[82]. Elle pourrait s'expliquer par la nature des chartes qui ne citent que certains types de biens et de ressources. Émerge ainsi une hypothèse : la rareté de ces deux « espèces » ne serait-elle pas imputable au fait qu'elles étaient exclues des dons et destinées à l'autoconsommation[83], tandis que le froment était [166] très valorisé et constituait un marqueur social ? La seconde phase irait de 900 à 1080 environ. Lors de celle-ci, c'est l'orge qui prend la tête des céréales les plus mentionnées, tandis qu'augmentent parallèlement les mentions de froment et d'avoine. L'épeautre et le seigle restent alors très faibles. Le troisième moment pourrait s'étendre de 1080 à 1150 environ. Au cours de celui-ci, l'orge stagne dans les chartes au profit du froment et de l'avoine qui explosent littéralement. On

---

est très forte, la proportion des types de céréales variant en fonction des écosystèmes régionaux et même locaux, mais aussi des stratégies de subsistance – ainsi que le montrent les inventaires et les fouilles.

[80] Cette approche ne va toutefois pas sans générer de biais, en particulier à cause de l'énorme variabilité régionale de l'Europe médiévale (cf. la note précédente). En l'occurrence, il faut conserver à l'esprit que cette lecture globale informe sur des tendances, mais que celles-ci « écrasent » des phénomènes complexes, liés aux différentes chronologies de la production documentaire et de la dynamique sociale, ainsi qu'à l'existence d'écosystèmes spécifiques. Une autre étude sur le même thème, cette fois à partir de corpus régionaux, offrirait probablement des résultats différents et donc complémentaires. Enfin, rappelons que ces tendances ne reflètent pas uniquement les pratiques agraires, mais aussi et essentiellement les discours dominants – les deux « lectures » se trouvant de fait à articuler.

[81] Tout en sachant que les données sont très difficiles à interpréter avant 800 à l'aide d'outils statistiques, à cause de la rareté relative des occurrences.

[82] Ces deux céréales étaient en effet particulièrement adaptées au système médiéval, parce qu'elles fournissaient des rendements plutôt stables (c'est-à-dire peu dépendants du sol et des aléas climatiques) et possédaient une bonne capacité de conservation : cf. A. Guerreau, « L'étude de l'économie médiévale… ».

[83] Le seigle est impropre à la production du pain blanc, ce qui pourrait expliquer cette absence de donation / codification. Son importance dans le système médiéval a pourtant été très tôt perçue : cf. C. Parain, « The Evolution of Agricultural Technique », M.M. Postan (dir.), *The Cambridge Economic History of Europe. I. The Agrarian Life of the Middle Ages*, t. 1, Cambridge, 1941 (repris en français dans *Outils, ethnies et développement historique*, Paris, 1979, p. 47-127, en particulier p. 100-101) ; F. Sigaut, « De l'écobuage au pain d'épice. Quelques questions sur l'histoire du seigle », J.-P. Devroey, J.-J. Van Mol, C. Billen (dir.), *Le seigle (Secale cereale), histoire et ethnologie…*, p. 211-250. Le cas de l'épeautre est plus subtil, car il peut permettre la création du pain blanc, mais seulement à condition de traiter les grains au préalable : voir J.-P. Devroey, J.-J. Van Mol (dir.), *L'épeautre, histoire et ethnologie…*

constate à ce moment que les mentions de seigle se développent, alors que celles de lieux de stockage pérennes progressent fortement (fig. 3, 4). On pourrait en conclure qu'à ce moment, l'intérêt pour la captation « régulée » des céréales évolue à la hausse au sein de l'aristocratie[84], mais aussi que la diversité des espèces en circulation augmente. Cette situation culmine lors de la dernière phase, entre 1150 et 1300, où l'épeautre émerge, en parallèle du développement du seigle. Ces céréales autrefois absentes des chartes sont maintenant bien présentes dans ces textes, à cause de la généralisation de l'écrit[85], mais aussi parce que l'essor du système et la seigneurialisation des rapports sociaux entraînent sans doute une plus forte codification de tout ce qui entoure les grains. Parallèlement à l'émergence de l'épeautre et du seigle dans les chartes, les mentions de froment et d'avoine explosent en effet littéralement. Enfin, le terme *bladum* (qui désigne le froment, mais qui est aussi très générique) se développe lui aussi fortement. Ce qui pourrait montrer que l'intérêt pour les grains est désormais généralisé et qu'une catégorie englobante s'est instaurée, en parallèle d'une circulation plus intense pour des espèces autrefois non-destinées aux transferts (épeautre, seigle), du moins ceux encadrés par les chartes.

L'information principale de ce graphique nous semble toutefois ailleurs : d'une part, dans la relative rareté des mentions de grains avant le milieu / la fin du XIe siècle ; d'autre part, dans la diversité des espèces mentionnées, qui augmente fortement une fois cette période dépassée. Autrement dit, le mouvement observé est strictement parallèle à celui [167] constaté pour les greniers. L'explosion des occurrences de céréales nous semble en effet imputable au renforcement de l'intérêt seigneurial pour la captation et la redistribution « polarisée » des grains. Ce basculement concorde avec l'idée que s'opère alors un passage du prélèvement fondé sur le travail (qui invisibilise donc le produit et son accumulation) à un prélèvement fondé sur la nature (qui invisibilise le rapport de domination, en faisant en revanche apparaître le produit, la codification du prélèvement, dans un temps encadré, et les lieux de stockage).

*Mesures, monnaies, redevances*

Peut-on aller plus loin et relever d'autres corrélations, entre le vocabulaire du stockage et certains ensembles lexicaux complémentaires ? Différents outils algorithmiques permettent aujourd'hui d'explorer les champs sémantiques de façon systématique, en révélant les proximités contextuelles, ainsi que l'évolution des associations entre différents termes[86]. Plusieurs analyses heuristiques ont donc été réalisées, dont seules les plus pertinentes sont ici présentées[87]. Le premier élément visible sur tous les graphiques est la proximité des contextes lexicaux des différents termes retenus – en particulier *horreum*, *grangia*, *granea*, *granarium* (fig. 6). En effet, *granica*, *granarium*, mais aussi *grangia* apparaissent comme des éléments sémantiquement proches d'*horreum* au XIII[e] siècle. Ce n'est pas le cas aux X[e]-XI[e] siècles, ce qui tendrait en outre à montrer qu'il y a une forme de rapprochement entre les différents lemmes (fig. 6a *vs* 6b ; 6c *vs* 6d). Le phénomène est aussi visible dans le champ sémantique de *granarium* au XIII[e] siècle, où apparaissent *horreum*, *grangia* et *granica* (fig. 7). Ce qui semble par ailleurs confirmer l'hypothèse selon laquelle ces termes forment un ensemble sémantique cohérent, sans toutefois qu'ils soient strictement synonymes. [168]

---

[84] Par captation « régulée », on entend un prélèvement fixe (dans la quantité et dans le temps), associé à un lieu précis où les débiteurs doivent verser le grain.

[85] Cette cause « techniciste » ne peut être considérée, à notre avis, comme première. Si la production des chartes augmente très fortement, c'est d'abord parce que les transferts de terres et les transactions autour de celles-ci explosent. En dehors des chartes, il est toutefois avéré que les pratiques scripturaires se développent énergiquement au cours de cette période : P. Bertrand, *Les écritures ordinaires. Sociologie d'un temps de révolution documentaire (entre royaume de France et Empire, 1250-1350)*, Paris, 2015.

---

[86] Il s'agit en particulier de la « sémantique distributionnelle », largement employée ici. Cette approche vise à dégager des contextes similaires pour différents lemmes, avec une hypothèse simple : les éléments linguistiques ayant des distributions similaires ont des significations similaires ; autrement dit, deux mots sont sémantiquement proches s'ils apparaissent dans des contextes similaires (synonymes, antonymes, cooccurrents significatifs, etc.). On trouve la base (théorique) de ce postulat dans J. Trier, *Der deutsche Wortschatz im Sinnbezirk des Verstandes : von den Anfängen bis zum Beginn des 13. Jahrhunderts*, Heidelberg, 1973. Les applications pratiques sont plus récentes, voir en particulier : S. Evert, « Distributional semantics in R with the Wordspace package », *Proceedings of COLING 2014, the 25th International Conference on Computational Linguistics: System Demonstrations*, Dublin, 2014, p. 110-114.

[87] Tous les mots sont connectés à une certaine échelle, chronologique ou de distance lexicale. La difficulté est donc de détecter les échelles d'associations les plus pertinentes, ce qui implique un certain nombre d'essais et de choix.

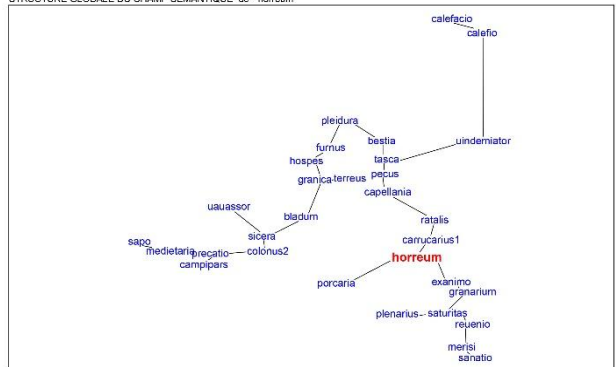
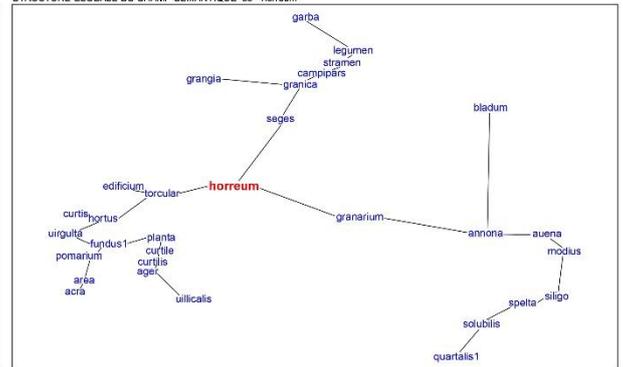
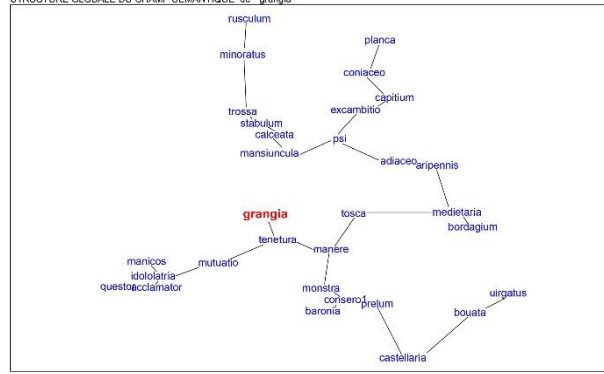
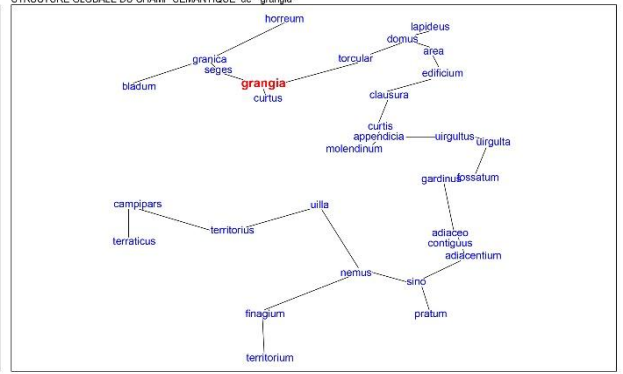

**Fig. 6.** Champs sémantiques dans les CEMA, méthode WSDSM : a) *horreum*, XI[e] siècle ; b) *horreum*, XIII[e] siècle ; c) *grangia*, XI[e] siècle ; d) *grangia*, XIII[e] siècle.

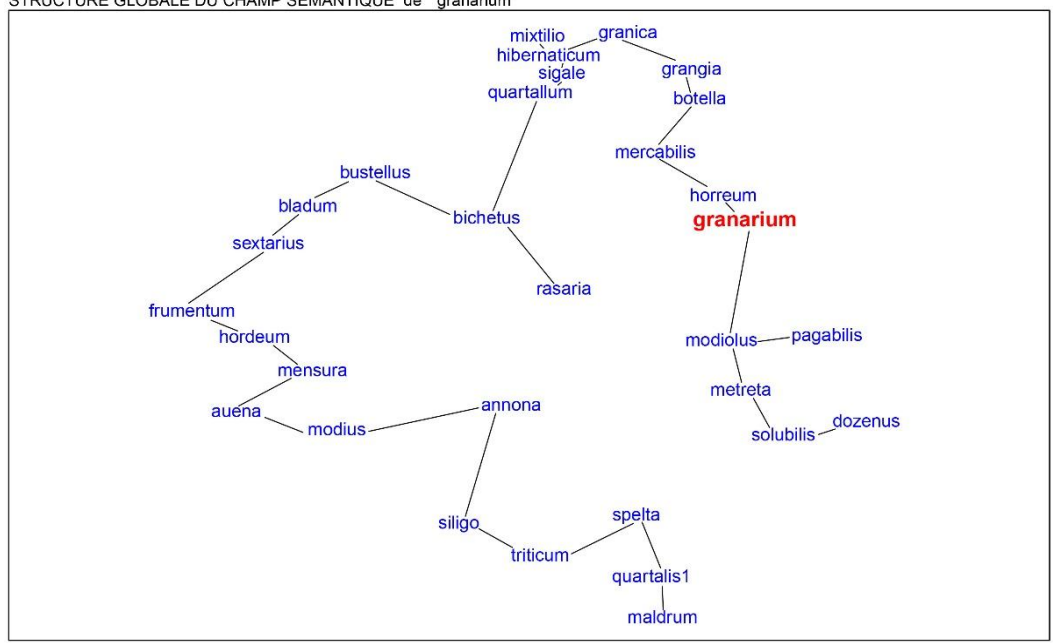

**Fig. 7.** Champ sémantique de *granarium* dans les CEMA (XIII[e] siècle), méthode WSDSM. On constate la proximité sémantique avec *horreum*, *grangia* et *granica*, qui apparaissent dans les cooccurrents de *granarium*.

[169] Parallèlement, à partir du milieu du XI[e] et de plus en plus nettement au XII[e] siècle, on constate un renforcement progressif des liens entre les lieux de stockage et la mesure – plus précisément la mesure des quantités de grains (fig. 8). Cela se manifeste par une présence renforcée de *mensura* dans les cooccurrents des lieux de stockage, mais aussi *modius*, *modiolus*, *bichetus*, *rasaria* ou *sextarius*. L'attraction entre les deux lexiques semble d'ailleurs si forte que les deux champs apparaissent en quasi-synchronie, se développant tous deux massivement aux XII[e]-XIII[e] siècles[88]. Les chartes mentionnent même des outils de mesures propres à certains greniers, à partir de la seconde moitié du XII[e] siècle[89]. C'est le cas en 1173 dans cet acte de l'évêque d'Évreux, Gilles, où se trouve réglé un conflit en faveur de l'abbaye Saint-Sauveur-le-Vicomte : « *abbas annuatim predicto Olivero solvet*

---

[88] Une analyse du champ sémantique de la « mesure » nous emmènerait évidemment trop loin. Voir A. Guerreau, « *Postscriptum. Mensura*, représentations du monde, structures sociales », *Histoire & Mesure*, vol. 16, 2001, p. 405-414 ; *id.*, « *Mensura* et *Metiri* dans la Vulgate », *Micrologus*, t. 19, 2011, p. 3-20 ; *id.*, « La mesure au Moyen Âge : quelques directions de recherche », *Mesure et histoire médiévale. XLIII[e] congrès de la SHMESP*, Paris, 2013, p. 17-40 ; W. Kula, *Les mesures et les hommes*, Paris, 1984.

[89] Une première recherche a permis de recenser plus de 30 occurrences, dont : « *perducendos ad mensuram grangie de Malregard* », dans un acte de l'évêque de Beauvais en 1173 (acte inédit transcrit par B.-M. Tock, à partir du cartulaire Beauvais, A.D. Oise, H 4650, t. 1, p. 11-12) ; « *in horreo de Holdrival ad mensuram granarii abbatie de Morneval* », à Morienval en 1179 (A. Peigné-Delacourt (éd.), *Cartulaire de l'abbaye de Morienval*, Senlis, 1879, n° 8).

*X^em quartaria frumenti ad mensuram orrei sui* »[90]. Ces éléments corroborent l'idée d'un plus grand contrôle autour des lieux de stockage, qui constituent un enjeu seigneurial majeur aux XI^e-XIII^e siècles. [170]

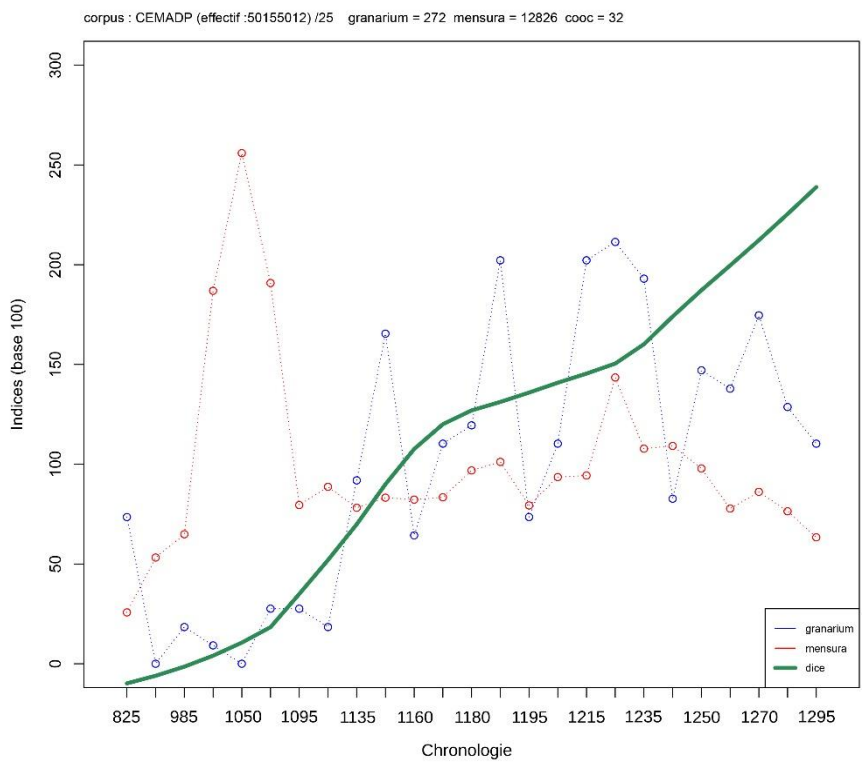

**Fig. 8.** Chronologie des associations (cooccurrences) entre *granarium* et *mensura* dans les CEMA (IX^e-XIII^e siècle). La ligne verte représente le taux d'association, calculé d'après un coefficient (Dice), à une distance de plus ou moins cinq mots.

Parmi les autres éléments notables autour des lieux de stockage, on trouve la question des dîmes et des redevances, sur laquelle les récents programmes dirigés par Michel Lauwers et Roland Viader ont renouvelé nos connaissances[91]. *Decima*, mais aussi *terraticus*, *campipars* se renforcent nettement dans le champ sémantique des lieux de stockage entre le XI^e et le XIII^e siècle (fig. 9), avec des centaines d'occurrences[92]. On constate [171] parallèlement que les greniers sont devenus des

---

[90] H. Müller (éd.), *Päpstliche Delegationsgerichtsbarkeit in der Normandie, 12. und frühes 13. Jahrhundert*, Bonn, 1997, t. 2, n° 27, p. 122.

[91] R. Viader (dir.), *La dîme dans l'Europe médiévale et moderne*, Toulouse, 2010 ; M. Lauwers (dir.), *La dîme, l'Église et la société féodale*, Turnhout, 2012. Ces liens ont aussi été évoqués dans L. Schneider, « De la fouille des villages abandonnés… », p. 146.

[92] Nous avons pu relever 413 associations entre *decima* et *horreum / granarium / grangia / granea* à plus ou moins cinq mots de distance (en incluant les textes non-datés ; soit 344 entre 825 et 1300). Aucune des mentions n'est toutefois antérieure à 1057 (« *quo domum et horreum ad suscipiendam et conservandam decimam fratres construere possint* », à l'abbaye Saint-Wandrille, dans M. Fauroux (éd.), *Recueil des actes des ducs de Normandie de 911 à 1066*, Caen, 1961, n° 190), et

lieux de tensions pour certaines communautés : *rapina*, tout comme *furtum* progressent parmi les cooccurrents principaux, avec des formules spécifiques[93], mais aussi les mentions d'incendies et de pillages déjà évoquées. Il est aussi probable que ce nouveau rapport aux lieux de stockage soit lié à l'essor des transactions monétaires, ainsi que le montre le développement des cooccurrences entre les termes désignant des lieux de stockage et *denarius*[94]. L'ensemble de ces termes renvoie globalement à un même phénomène (néanmoins complexe) : celui de la généralisation de la codification des redevances, dont la temporalité, le contenu et la quantité sont largement mis par écrit (et donc plus ou moins « fixés ») à partir des XI$^e$-XII$^e$ siècles[95]. C'est dans cette dynamique socio-spatiale que s'inscrit le développement des lieux de stockage identifiés et bien contrôlés par des individus précis, bâtis en bois ou en pierre. [172]

---

l'unique occurrence précoce provient d'un pseudo-acte de Louis le Pieux pour Saint-Zénon de Vérone (T. Kölzer (éd.), *Die urkunden Ludwigs des Frommen*, Wiesbaden, 2016, n° 74). Les associations des lieux de stockage avec *terraticus* et *campipars* sont certes moins nombreuses (59 mentions relevées) mais ces lemmes sont beaucoup moins fréquents que *decima*. Au demeurant, cette évolution indique un lien fort entre la question des redevances et la création de lieux de stockages dédiés à partir du second tiers du XI$^e$ siècle.

[93] En particulier dans les formulaires pontificaux dédiés aux confirmations. Deux exemples parmi plusieurs centaines : « *Paci quoque et tranquillitati vestre paterna sollicitudine providentes auctoritate apostolica prohibemus ut nullus infra clausuram locorum vestrorum sive grangiarum vestrarum violentiam vel rapinam facere vel hominem illicite capere audeat* », en 1147, dans une bulle d'Eugène III pour l'abbaye de Beaupré (acte inédit, transcrit par B.-M. Tock à partir du cartulaire Paris, BNF, lat. 9973, fol. 127r) ; « *prohibemus ut nullus infra fines locorum vestrorum seu grangiarum vestrarum furtum facere ignem apponere seu hominem capere vel interficere* », en 1184, dans un acte de Lucius III pour Chichester (P.M. Hoskin (éd.), *English Episcopal Acta (22). Chichester 1215-1253*, Oxford, 2001).

[94] Sur la question des monnaies, voir en dernier lieu : L. Kuchenbuch, *Versilberte Verhältnisse : der Denar in seiner ersten Epoche (700-1000)*, Göttingen, 2016.

[95] Ce qui n'implique en rien une forme d'uniformisation. Tout au contraire, on peut penser que le système médiéval aurait par essence rejeté une telle tentative, puisqu'elle aurait conduit à une « dépolarisation », par la possibilité de permuter les terres, mais aussi le surplus qui en était extrait.

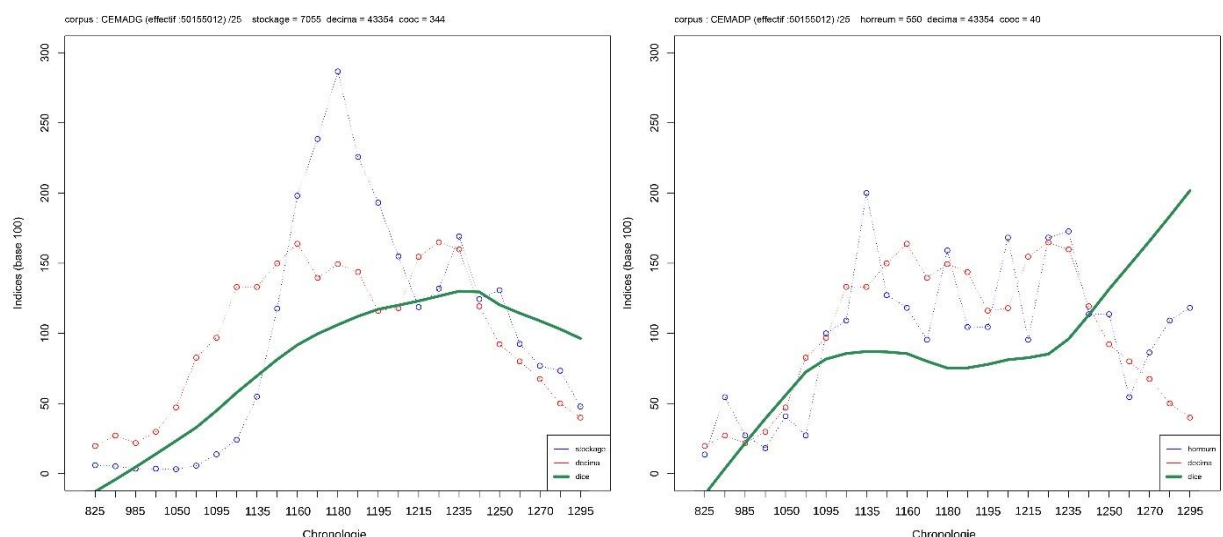

**Fig. 9.** Chronologie des cooccurrences dans les CEMA (IX<sup>e</sup>-XIII<sup>e</sup> siècle), entre : a) *horreum/grangia/granarium/granea* et *decima* ; b) *horreum* et *decima*. La ligne verte représente le taux d'association, calculé d'après un coefficient (Dice), à une distance de plus ou moins cinq mots.

*Le grenier : un espace-temps bon à penser*

On a pu par ailleurs constater un renforcement très sensible des liens entre les greniers et les champs sémantiques de l'espace, d'une part, et du temps, d'autre part, au cours des XI<sup>e</sup>-XIII<sup>e</sup> siècles. La dimension temporelle apparaît évidemment articulée aux redevances et aux cycles que celles-ci impliquent : *tempus*, *annus*, etc. (fig. 10a)[96]. Dans le cas de l'espace, les manifestations sont multiples : on note l'émergence d'un assez grand nombre de toponymes formés à partir des noms de greniers, aux X<sup>e</sup>-XIII<sup>e</sup> siècles[97]. Par ailleurs, ces derniers sont de plus en plus associés à *locus* (fig. 10b). Un tel indice laisse à penser que le grenier devient un lieu spécifique, polarisant et donc fixe[98] – mais aussi l'occasion d'un dialogue potentiel entre dominants et producteurs. *Communis* progresse d'ailleurs au même moment dans l'entourage lexical d'*horreum*, de *granarium* et de *granea*. Enfin, les fonctions liées à ces lieux se développent aussi : *custos* [173] *horrei* ou encore *grangiarius* (193 mentions à partir de 1150) – ce qui corrobore l'idée d'une forme d'institutionnalisation des greniers.

---

[96] L'articulation entre temps et espace, dans les chartes et en général, reste un domaine peu exploré en médiévistique.
[97] Quelques exemples de toponymes formés sur ces lemmes : « *villa Granarium* », « *que vocatur Horreum* », « *fiscum de Granario* », « *curte qui dicitur Orrea* », « *Horreensi cenobio* », « *rivo Granario* », « *que vulgo dicitur Granaroy* », etc.
[98] A. Guerreau, « Il significato dei luoghi nell'Occidente medievale : struttura e dinamica di uno *spazio* specifico », E. Castelnuovo, G. Sergi (dir.), *Arti e Storia nel Medioevo*, v.1., Torino, 2002, p. 201-239.

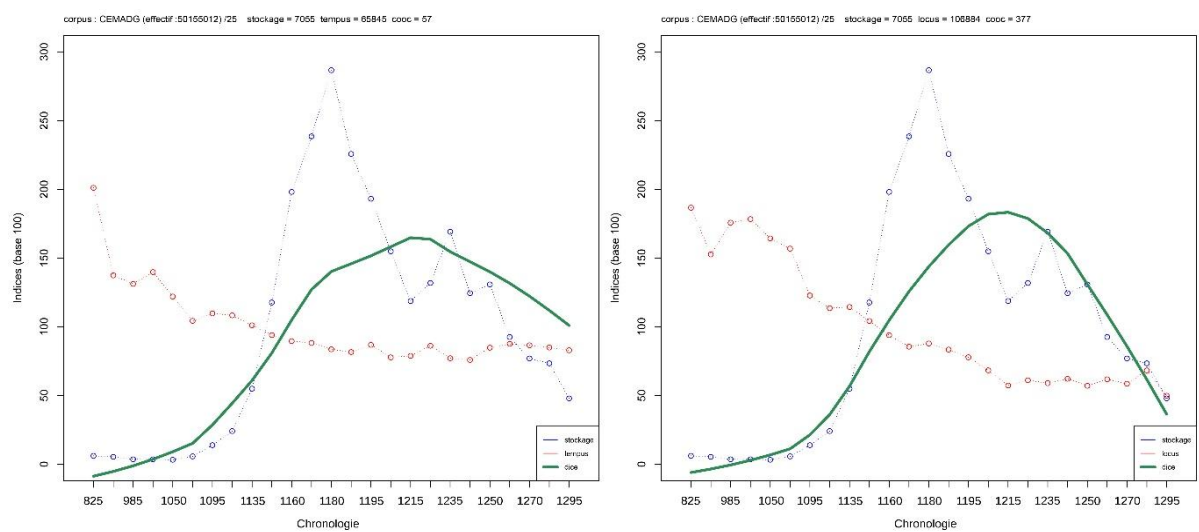

**Fig. 10.** Chronologie des cooccurrences dans les CEMA (IX[e]-XIII[e] siècle), entre : a) *horreum / grangia / granarium / granea* et *tempus* ; b) *horreum / grangia / granarium / granea* et *locus*. La ligne verte représente le taux d'association, calculé d'après un coefficient (Dice), à une distance de plus ou moins cinq mots.

L'analyse numérique des contextes lexicaux permet encore de relever différentes évolutions dans l'environnement des lieux de stockage. Jusqu'au XI[e] siècle, une grande diversité paraît prévaloir, accompagnée d'une faiblesse dans la localisation. Les greniers semblent alors s'insérer dans un tissu rural moins dense, autour du *campus* et de l'*ager*, parfois même incertain dans sa structuration[99]. À partir du XII[e] siècle, un certain nombre de réserves apparaît dans les chartes en proximité de forêts – à travers des cooccurrents tels que *nemus*, *forestis* et *silva*[100]. Dans d'autres séries de mentions cependant, les liens entre les greniers et d'autres édifices [174] se renforcent : c'est notable pour *domus*[101]. Progressivement, on voit apparaître des greniers à proximité d'autres bâtiments, moulins (*molendinum*), pressoirs (*torcular*) et églises[102]. Les occurrences associant ces lieux et le lemme

---

[99] C. Wickham, *Framing the Early Middle Ages. Europe and the Mediterranean, 400-800*, Oxford, 2005, p. 516-518. Cette relative fluidité du tissu rural ne doit pas pour autant conduire à l'idée que des principes intellectuels globalement cohérents ne sous-tendaient pas l'organisation spatiale et la production.

[100] La seule occurrence associant un terme « forestier » à un stockage en grenier, antérieure à 1098, provient d'un document faux / interpolé (P. Kehr (éd.), *Die Urkunden Karls III...*, n° 84 – 883). À partir de cette date, nous avons pu relever 330 mentions, essentiellement aux XII[e] et XIII[e] siècles. Comme le phénomène est chronologiquement concomitant à la « monumentalisation » des greniers, on peut aussi faire l'hypothèse qu'outre un rapprochement avec les espaces boisés, ce mouvement est aussi lié à l'utilisation du bois pour la construction des lieux de stockage.

[101] Au total, 579 mentions associant *domus* aux lieux de stockage ont été repérées, principalement à partir de 1050.

[102] Deux exemples parmi beaucoup d'autres : « *cum omnibus terris et pratis que ad ipsam carrucam pertinent et horreum ante ecclesiam et domum que ad levam ipsius horrei sita est* », à Conques en 1107 (G. Desjardins (dir.), *Cartulaire de l'abbaye de Conques en Rouergue*, Paris, 1879, n° 485) ; « *Dedit etiam domum suam et horreum iuxta ecclesiam* », à Saint-Mihiel en 1121 (A. Lesort (dir.), *Chroniques et chartes de l'abbaye de Saint-Mihiel*, Paris, 1909-1912, n° 62).

*ecclesia* se développent en effet fortement entre la fin du XIe et la fin du XIIe siècle – tout en sachant que ce lemme peut bien entendu désigner des édifices, mais aussi l'institution elle-même[103]. Le phénomène est d'autant plus remarquable que les associations entre *horreum / grangia / granarium / granea* et *castrum* sont plutôt rares[104]. Tout ceci laissant penser, à la fois à travers les dîmes et les mentions d'édifices, que l'institution ecclésiale a joué un rôle majeur dans ce développement des greniers.

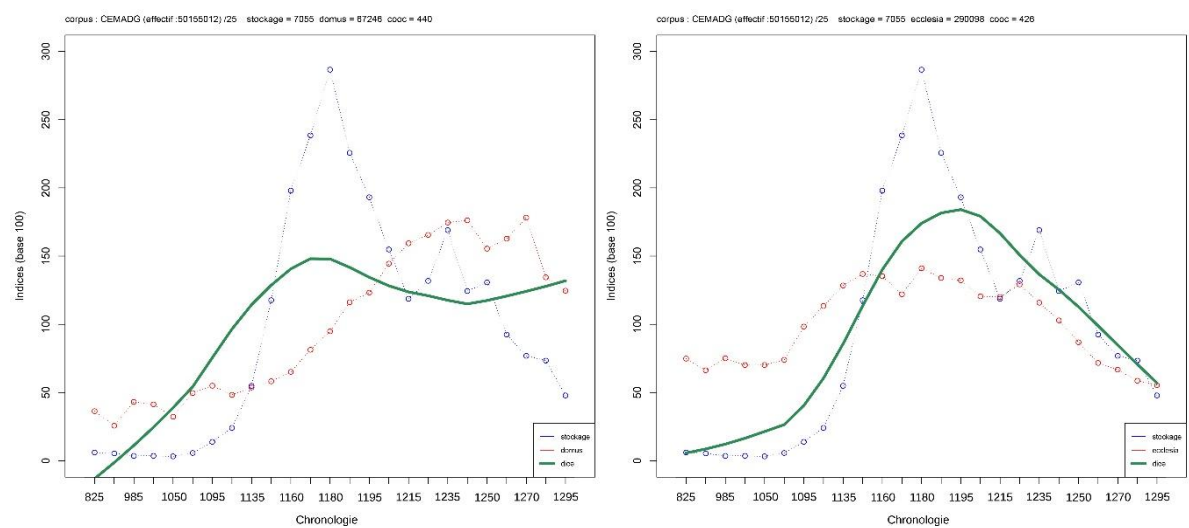

**Fig. 11.** Chronologie des cooccurrences dans les CEMA (IXe-XIIIe siècle), entre : a) *horreum / grangia / granarium / granea* et *domus* ; b) *horreum / grangia / granarium / granea* et *ecclesia*. La ligne verte représente le taux d'association, calculé d'après un coefficient (Dice), à une distance de plus ou moins cinq mots.

[175] Il a par ailleurs été possible de constater que ces lieux étaient très nettement liés, au moins à partir du XIIe siècle, à un certain nombre de verbes impliquant des actions et des déambulations (*affero*, *colligo*, *recondo*, *adduco*, *cooperio*, *duco*, *conduco*, *vehiculo,* etc.)[105]. On se rend ainsi au grenier, où l'on stocke une partie de sa récolte, à destination de la communauté mais surtout du

---

[103] La question n'est donc pas seulement l'association entre les lieux de stockage et l'édifice ecclésial, mais aussi entre ces greniers et l'Église en tant que structure englobante d'organisation sociale.

[104] Quelques dizaines seulement sur l'ensemble des CEMA. Cela va sans dire : de telles réserves existaient nécessairement (cf. J.-M. Poisson, « Espaces et modes de stockage… »). Le fait qu'elles soient absentes de la documentation diplomatique induit différentes hypothèses : soit leur présence était si évidente qu'il n'était pas nécessaire de les mentionner ; soit les greniers mentionnés dans les chartes relèvent d'autres logiques (communautaires, seigneuriales, ecclésiales).

[105] Par exemple : « *et decimam recipiet sed monacus grangie sicut suas conducet ad horreum suum* », à La Sauve-Majeure en 1096 (C. Higounet, A. Higounet-Nadal, N. Peña (éd.), *Grand cartulaire de La Sauve-Majeure*, Bordeaux, 1996, n° 1410) ; « *Decima quoque et vinearum et agrorum que cellerarius propriis bobus vel sumptibus excoluerit in sacriste horrea debet recondi* », à Auch en 1140 (C. Lacave La Plagne Barris (éd.), *Cartulaires du chapitre de l'église métropolitaine Sainte-Marie d'Auch. Cartulaire noir*, Paris-Auch, 1899, n° 88).

s/Seigneur[106]. Cette déambulation liée aux redevances, dans un espace-temps encadré, matérialise visuellement la puissance seigneuriale – à l'inverse des fosses-silos du haut Moyen Âge, qui n'entraînaient *a priori* pas de ritualisation particulière, ni de déplacement signifiant.

*Variabilité géographique du phénomène*

Avant d'apporter quelques éléments conclusifs, il convient cependant de nuancer certaines des observations réalisées jusqu'ici. Les mentions de lieux de stockage sont en effet beaucoup plus présentes dans les corpus du nord de l'Europe – disons à partir de la région Rhône-Alpes ou de la Bourgogne, pour l'actuelle France. Cette répartition très nette pourrait en partie être imputable à des conditions géo-climatiques différentes – bien que François Sigaut ait autrefois refusé ce type d'argument plus ou moins déterministe[107]. L'ensilage produisait probablement de meilleurs résultats dans les espaces méridionaux en général[108]. Cela expliquerait-il les différences observées entre nord et sud, ces dernières zones ayant eu moins besoin de recourir aux greniers en élévation, en tout cas moins rapidement ? Il se pourrait en effet que cette meilleure conservation ait non seulement favorisé la mise en réserve enterrée dans le haut Moyen Âge, mais aussi qu'elle ait conduit à conserver ces techniques plus longtemps qu'au nord – où les résultats n'étaient probablement pas optimaux.

S'il ne s'agit là que d'une hypothèse, elle nous semble néanmoins intéressante : pour l'anthropologue Alain Testart, les greniers se trouvent en [176] partie à l'origine de l'inégalité dans les sociétés humaines[109]. Ne faudrait-il pas voir là une des causes multiples de la variabilité du système médiéval, à travers l'inégalité du processus de seigneurialisation, entre nord et sud ? Cela reviendrait à revaloriser le grenier en tant qu'objet de civilisation, réinventé pleinement par l'Europe médiévale, ainsi qu'en témoignent les textes narratifs et le développement progressif d'un *horreum celestis*, analogon spirituel des infrastructures de mise en réserve. Ne pouvant mener cette enquête ici, nous proposons de la conduire dans une étude ultérieure, qui sera consacrée à la variabilité géographique du stockage céréalier.

---

[106] Sur les liens entre organisation spatiale, domination et systèmes processionnaires, voir L. Kuchenbuch, J. Morsel, D. Scheler, « La construction processionnelle de l'espace communautaire », D. Boisseuil, P. Chastang, L. Feller, J. Morsel (dir.), *Écriture de l'espace social. Mélanges d'histoire médiévale offerts à Monique Bourin*, Paris, 2010, p. 139-182.
[107] F. Sigaut, « La technologie de l'agriculture. Terrain de rencontre entre agronomes et ethnologues », *Études rurales*, vol. 59, 1975, p. 103-111.
[108] Voir par exemple la note 56 sur *fovea*.
[109] Voir les références mentionnées en introduction.

*
* *

**Éléments conclusifs**

L'analyse systématique des occurrences des lieux de stockage dans les CEMA a permis d'observer un changement majeur au cours des XIe-XIIIe siècles, tant quantitatif que sémantique. Pour le haut Moyen Âge, ces textes ne mentionnent pas la présence de fosses-silos, pourtant bien observées en archéologie. Avant la seconde moitié du IXe siècle, les rares occurrences de greniers proviennent en effet majoritairement de documents à la limite de la typologie des chartes (chartes-inventaires, polyptyques, listes). Ici comme ailleurs, les corpus textuels et archéologiques, loin de s'opposer, se complètent donc, parce qu'ils offrent différents angles de vue sur un même objet. Les fosses-silos nous apparaissent ainsi comme l'expression d'une stratégie de stockage moins hiérarchisée, peut-être moins régulée – ce qui n'exclut pas une logique communautaire, bien au contraire[110]. Une hypothèse de travail est que les dominants se focalisent alors sur ce qu'ils obtiennent (le produit), mais pas sur la production, ni sur le stockage de cette dernière.

L'explosion du nombre de mentions aux XIIe-XIIIe siècles traduit ainsi, nous semble-t-il, un double phénomène : d'une part, une augmentation probable de ces structures ; d'autre part, aussi, une évolution du regard [177] porté sur celles-ci. Les greniers s'insèrent alors très fortement dans la logique des redevances, de la centralisation, donc de la seigneurialisation, à laquelle l'Église participe fortement. Sémantiquement, ces réserves se lient à d'autres champs – le temps, l'espace, les grains, la mesure, les communs, les édifices, etc. – qui montrent que le grenier devient un lieu polarisant. Celui-ci participe ainsi à la spatialisation observable dans l'Europe médiévale au cours des Xe-XIIIe siècles, à travers des déplacements vers celui-ci, et à une organisation nouvelle de redistribution de la production, dans un espace-temps régulé par les élites. Or, si l'on repense à la première partie de cette analyse, c'est précisément au cours de cette période que se développent les préambules évoquant les greniers célestes. Il y a donc, parallèlement, une réflexion sur le grenier en tant qu'infrastructure polarisante, manifestation et incarnation du pouvoir seigneurial, et le développement d'une pensée de la mise en réserve.

---

[110] On retrouverait peut-être là certaines hypothèses de Chris Wickham sur le mode de production du haut Moyen Âge (« *peasant-based* », sans que cela implique la théorie du « mode de production paysan » au sens strict). Voir C. Wickham, « Problems of Comparing Rural Societies in Early Medieval Western Europe », *Transactions of the Royal Historical Society*, 1992, vol. 2, p. 221-246, ainsi que le rapport de M. Lauwers dans le présent volume.